\documentclass[aps,onecolumn,tightenlines,amsmath,amssymb,11pt,superscriptaddress,nofootinbib]{revtex4}

\usepackage{graphicx}
\usepackage{amsmath,amssymb,amsfonts,amsthm,stmaryrd,mathtools,mathbbol,bm,physics,tensor}
\usepackage{soul}
\usepackage{xcolor}
\usepackage{tikz}
\allowdisplaybreaks[1]
\usepackage[bookmarks,linktocpage, colorlinks=true, plainpages = false, citecolor = treegreen,  linkcolor=darkblue, urlcolor = darkblue, filecolor = blue]{hyperref} 

\definecolor{tealgreen}{rgb}{0.0, 0.5, 1.0}
\definecolor{darkblue}{rgb}{0., 0.4, 0.8}
\definecolor{cadmiumred}{rgb}{1., 0., 0.22}
\definecolor{treegreen}{rgb}{0., 0.7, 0.3}
\usepackage{pifont}

\def\be#1\ee{\begin{align}#1\end{align}}

\def\ba{\begin{eqnarray}}
\def\ea{\end{eqnarray}}
\def\nn{\nonumber}

\usepackage[normalem]{ulem}

\begin{document}

\title{Formation of extremal regular black holes
}

\author{Johanna Borissova}
\email{j.borissova@imperial.ac.uk}
\affiliation{Abdus Salam Centre for Theoretical Physics, Imperial College London, London SW7 2AZ, United Kingdom}

\begin{abstract}
\bigskip
{\sc Abstract:} Static and slowly evolving regular black holes with a non-extremal inner horizon are generally expected to suffer from mass inflation. Here we consider the scenario of asymptotic gravitational collapse into extremal regular black holes.
To that end, we first construct geometric models of static spherically symmetric single-horizon extremal and double-horizon inner-extremal regular black holes for generic black hole mass as the only dimensionful scale. These spacetimes may be interpreted as non-fine-tuned vacuum solutions of modified gravitational theories
defined implicitly by the requirement that their spherical reduction yields an integrable two-dimensional Horndeski theory. As such, these so-called general quasi-topological gravities admit exact Vaidya solutions in which the mass becomes a time-dependent function. We use this effectively two-dimensional second-order dynamical framework to model the asymptotic formation of extremal regular black holes without a violation of energy conditions. The latter is illustrated explicitly by the equivalence between the strong energy condition for generalised Vaidya solutions of general relativity, and 
the kinematic timelike convergence condition reformulated dynamically as an onshell condition on the theory-dependent functions characterising the spherical reduction and correspondingly Vaidya solutions of a general quasi-topological gravity. The defocusing of geodesics necessary for regular black holes is thus triggered by the modified gravitational dynamics rather than by the addition of exotic matter degrees of freedom.
\\

\noindent Comments: Invited contribution submitted to the {\it Focus on Regular Black Holes} collection to appear in {\it Classical and Quantum Gravity}.
\bigskip
\noindent

\end{abstract}

\maketitle

\tableofcontents

\bigskip\bigskip

\section{Introduction}\label{Sec:Introduction}

Regular black holes~\cite{Bardeen:1968bh,Hayward:2005gi,Dymnikova:1992ux,Fan:2016hvf} are candidate geometries in which the spacetime singularity of vacuum black holes in general relativity is replaced by a non-singular core. Such a singularity regularisation in general relativity requires the coupling to matter violating the strong energy condition (SEC)~\cite{Zaslavskii:2010qz,Wang:2026jvo,Wang:2026sqr}. An example are regular black holes sourced by non-linear electromagnetic fields~\cite{Ayon-Beato:1998hmi,Ayon-Beato:1999qin,Ayon-Beato:1999kuh,Ayon-Beato:2000mjt,Bronnikov:2000vy,Dymnikova:2004zc,Bronnikov:2022ofk,Balart:2014cga,Fan:2016hvf,Rodrigues:2018bdc,AraujoFilho:2026sna}. This is because the SEC is equivalent to the geometric timelike convergence condition (TCC) when the Einstein equations hold, while the latter is one of the key assumptions of the Penrose-Hawking 1970 singularity theorem~\cite{Hawking:1970zqf,Hawking:1973uf} and a purely kinematic condition on the Ricci tensor. The TCC ensures the focusing of causal geodesics and is generally violated in regular black hole spacetimes~\cite{Borissova:2025hmj}. It contains the geometric null convergence condition (NCC) as a limiting case and  one of the main assumptions of the Penrose 1965 singularity theorem~\cite{Penrose:1964wq,Hawking:1973uf}. The NCC is equivalent to the null energy condition (NEC) in general relativity. However, this earlier theorem, while featuring a weaker convergence condition, imposes the stronger causality condition of global hyperbolicity which is violated in static regular black holes with a Cauchy horizon. Therefore, while such kinds of regular black hole geometries do not necessarily violate the NCC, they in any case circumvent the Penrose theorem by violating the assumption of global hyperbolicity~\cite{Borissova:2025msp}. This conclusion in general does not apply to dynamical regular black holes, and in particular transient regular black holes necessarily violate the NCC~\cite{Borissova:2025msp,Borissova:2026ohj}. Ultimately, the NCC and TCC represent the relevant geometric conditions required in applications of the singularity theorems, while the NEC and SEC are only meaningful, in what concerns the application of these theorems, provided the Einstein equations hold~\footnote{Or else matter energy conditions in modified gravity theories would have to be converted, via the convergence conditions, into conditions which hold dynamically onshell.} --- 
an assumption about the gravitational dynamics which is generally expected to break down at high energies.  \\

In a more complete perspective on classical and quantum gravity, one may expect the regularisation of black hole singularities to result from  higher-derivative and non-local
terms in the gravitational effective action.
In this context, a few largely model-independent observations about the significance of spacetime singularities apply to higher-derivative gravitational actions. 
On the one hand, considering a path integral for quantum gravity with a microscopic action built from all local operators compatible with diffeomorphism symmetry, i.e., in particular involving  higher-curvature invariants beyond the Ricci scalar, singular black hole and cosmological spacetimes may be effectively offshell suppressed due to large gradients of the action between neighboring configurations resulting in destructive interference~\cite{Lehners:2019ibe,Borissova:2020knn,Borissova:2023kzq,Borissova:2024hkc,Giacchini:2021pmr,Chojnacki:2021ves}. On the other hand, the central question then still remains whether onshell configurations of the effective action, obtained by integrating out quantum fluctuations from the gravitational path integral, are regular. Addressing this question is challenging not least because the ultimate gravitational effective action is unknown, and only few observational constraints restricting the subleading terms in a derivative expansion are available~\cite{Calmet:2008tn,Hoyle:2004cw}. Nevertheless, there are several indications pointing towards the possibility that higher-derivative effective actions may indeed give rise to regular black holes. For instance, local regular series solutions at a would-be black hole core have been found in quadratic gravity~\cite{Stelle:1977ry,Lu:2015cqa,Lu:2015psa,Podolsky:2018pfe,Podolsky:2019gro} and quasi-local modifications thereof~\cite{Borissova:2025nvj}, as well as in six-derivative gravity~\cite{Giacchini:2024exc,Giacchini:2025gzw}.~\footnote{Such local series analyses a priori do not imply that there exist global regular black hole solutions.} Moreover, qualitative insights from asymptotically safe quantum gravity~\cite{Bonanno:2000ep,Reuter:2003ca,Bonanno:2006eu,Falls:2010he,Torres:2014gta,Koch:2013owa,Kofinas:2015sna,Bonanno:2016dyv,Bonanno:2017zen,Bonanno:2017kta,Pawlowski:2018swz,Adeifeoba:2018ydh,Held:2019xde,Platania:2019kyx,Borissova:2022mgd,Borissova:2022jqj,Bonanno:2023rzk,Bonanno:2024paf,Bonanno:2024wvb} suggest that higher derivatives and non-localities may indeed contribute to black hole singularity regularisation.

The arguably most compelling evidence for singularity regularisation induced by higher-derivative terms in the gravitational action stems from vacuum regular black holes in polynomial curvature quasi-topological gravities~\cite{Oliva:2010eb,Myers:2010jv,Dehghani:2011vu,Cisterna:2017umf,Bueno:2019ltp,Bueno:2019ycr, Bueno:2022res,Moreno:2023rfl}  in five and higher dimensions~\cite{Bueno:2024dgm}. The relevant densities provide a basis for a systematic curvature expansion of the action, whereby an infinite tower of curvatures is required to regularise the Schwarzschild singularity. Such exact vacuum regular black hole solutions supported by standard effective field theory in four spacetime dimensions are to our knowledge not known.~\footnote{However, see~e.g.~\cite{Buoninfante:2022ild,dePaulaNetto:2023cjw,Cadoni:2026cfi} for tentative regular black hole solutions in approximated  field equations within the non-local form-factor approach, and e.g.~\cite{Knorr:2022kqp} for a general discussion of conditions for regular black holes to arise from an action principle based on large-distance approximated field equations and solutions.} Nevertheless,  one may construct non-polynomial gravitational actions in four dimensions whose spherical reduction yields the relevant classes of two-dimensional Horndeski theories obtained from the reduction of higher-dimensional polynomial curvature quasi-topological gravities~\cite{Bueno:2025zaj,Borissova:2026wmn,Borissova:2026krh}. See also~\cite{Colleaux:2015yta,Colleaux:2019ckh,Colleaux:2026qew} for related non-polynomial Lagrangian approaches to regular black holes in four dimensions. In fact, generic two-dimensional Horndeski theories can be lifted to $d\geq 4$ dimensional gravities~\cite{Borissova:2026krh,Borissova:2026wmn,Colleaux:2017ibe,Colleaux:2019ckh}, such that realising the degrees of freedom of large classes of spherically symmetric regular black hole metrics as solutions to the equations of motion of two-dimensional Horndeski theories~\cite{Carballo-Rubio:2025ntd,Boyanov:2025pes,Kunstatter:2015vxa} amounts to realising these spacetimes as vacuum solutions to appropriate higher-dimensional gravitational theories. See in detail the discussion in~\cite{Borissova:2026wmn,Borissova:2026krh} and~e.g.~\cite{Borissova:2026rbi} for a summary. When lifting two-dimensional integrable Horndeski theories to higher-dimensional gravities, the resulting theories satisfy a generalised version of the Birkhoff theorem for polynomial curvature quasi-topological gravities~\cite{Bueno:2025qjk}, as discussed in~\cite{Borissova:2026wmn,Borissova:2026krh} --- which motivates  terming these gravitational theories {\it general quasi-topological gravities}. We will use this terminology throughout, i.e., the gravitational theories here referred to as general quasi-topological gravities are identified implicitly based on their spherical reduction, which are two-dimensional integrable Horndeski theories. As such, these theories are associated with an extension of the relevant space of spherically reduced actions to include generic two-dimensional integrable Horndeski theories~\cite{Borissova:2026krh}.~\footnote{The recently considered non-local modifications of higher-dimensional polynomial curvature quasi-topological gravities~\cite{Bueno:2026oyg} do not belong to the class of general quasi-topological gravities in the sense referred to here, as their spherical reduction produces higher-order equations of motion. Moreover, the notion of a general quasi-topological gravity used here should not be confused with the notion of a generalised quasi-topological gravity~\cite{Bueno:2019ltp,Bueno:2019ycr, Bueno:2022res,Moreno:2023rfl}. The latter refers to polynomial higher-curvature extensions of general relativity admitting static spherically symmetric vacuum solutions satisfying $g_{tt}g_{rr} = -1$, whereby the equation determining $g_{tt}=-f(r)$  is algebraic only if the generalised quasi-topological gravity is a polynomial curvature quasi-topological gravity. Otherwise the spherical reduction of such theories does not yield a two-dimensional integrable Horndeski theory.} 

Concretely, the associated Birkhoff theorem states that generic spherically symmetric solutions in vacuum are static and can be gauge-fixed to satisfy $g_{tt}g_{rr}=-1$ in Schwarzschild gauge, whereby $g_{tt}=-f(r)$ is entirely determined by an algebraic equation~\cite{Borissova:2026krh,Borissova:2026wmn} --- see e.g.~\cite{Bueno:2025qjk} for the particular case of higher-dimensional polynomial curvature quasi-topological gravities.
A related more general Birkhoff theorem in the context of generic two-dimensional Horndeski theories is stated in~\cite{Carballo-Rubio:2025ntd}. 

If the generating function characterising the spherical reduction of a quasi-topological gravity admits an expansion with an infrared limit corresponding to general relativity, then  for asymptotically  flat or anti-de Sitter black hole solutions the integrated equation of motion determining $f$ involves the Arnowitt-Deser-Misner (ADM) mass~\cite{Arnowitt:1959ah,Arnowitt:1962hi} as an integration constant. A given such spacetime is thus generated as a vacuum solution for all values of the mass $M$ --- differently e.g.~from sourcing such regular black holes in general relativity coupled to non-linear electrodynamics~\cite{Ayon-Beato:1998hmi,Ayon-Beato:1999qin,Ayon-Beato:1999kuh,Ayon-Beato:2000mjt,Bronnikov:2000vy,Dymnikova:2004zc,Bronnikov:2022ofk,Balart:2014cga,Fan:2016hvf,Rodrigues:2018bdc,AraujoFilho:2026sna}. This same observation can be used in reverse to reconstruct the degrees of freedom of such spacetimes as solutions to the equations of motion of two-dimensional integrable Horndeski theories~\cite{Carballo-Rubio:2025ntd,Boyanov:2025pes}, and thereby as vacuum solutions to higher-dimensional general quasi-topological gravities~\cite{Borissova:2026wmn,Borissova:2026krh}.
Our goal here will be to exploit the extended spherically symmetric solution space of general quasi-topological gravities~\cite{Borissova:2026krh}
 compared to the one of $d\geq 5$ polynomial curvature quasi-topological gravities~\cite{Bueno:2025qjk},  
 to interpret static spherically symmetric extremal regular black holes with mass parameter $M$ as the only dimensionful scale as vacuum solutions for generic black hole mass, without requiring a fine-tuning between $M$ and potential coupling constants of the theory.~\footnote{The construction of extremal regular black holes in polynomial curvature quasi-topological gravities without fine-tuning is obstructed by the fact that in this case the algebraic equation involving $M$ must depend on $r$ and $f$ in a constrained way through the Riemann scalar $\psi(r,f) = \frac{1-f}{r^2}$. See e.g.~\cite{DiFilippo:2024mwm} for a construction of fine-tuned double-horizon inner-extremal regular black holes in higher-dimensional polynomial curvature quasi-topological gravities, and e.g.~\cite{Liu:2026ltw} for a construction based on a four-dimensional analogue algebraic equation for $f$ in the context of polymerised vacuum solutions inspired by loop quantum gravity, which requires a similar fine-tuning. Taking into account generic classes of integrable two-dimensional Horndeski theories, which span the full extended space of spherically reduced general quasi-topological gravities, this issue does not occur~\cite{Borissova:2026krh}.}\\ 

The main motivation for considering extremal regular black holes stems from mass inflation expected in black hole spacetimes with a non-extremal inner horizon, i.e., an inner horizon with non-vanishing surface gravity which triggers an exponential growth of gravitational energy when perturbed by ingoing and outgoing fluxes, as originally discussed in the context of the Reissner-Nordstr\"om solution in general relativity~\cite{Poisson:1989zz,Poisson:1990eh,Ori:1991zz}. This same effect is expected to render static or slowly evolving non-extremal regular black holes classically instable~\cite{Brown:2011tv,Carballo-Rubio:2018pmi,Bertipagani:2020awe,Carballo-Rubio:2021bpr,DiFilippo:2022qkl,Carballo-Rubio:2022pzu,Carballo-Rubio:2024dca}, even though the details of the derivation leading to   such a conclusion are likely in general to depend on the particular gravitational theory under consideration.
Regardless of whether mass inflation in general quasi-topological gravities is a potentially concerning phenomenon or not --- see~\cite{Frolov:2026rcm,DiFilippo:2026jpv} addressing this question in higher-dimensional polynomial curvature quasi-topological gravities, and also~\cite{Droz:1994aj,Balbinot:1994ee,Chan:1994tb,Cai:1995nt,Frolov:2006is} in associated closely related two-dimensional dilaton theories --- our analysis is in parts tailored so as to not require an answer to this question, with the primary intention being, ideally, to not require asking this question in the first place. \\

Concretly, we will first construct geometric models of four-dimensional static spherically symmetric single-horizon extremal regular black holes and double-horizon inner-extremal regular black holes~\cite{Carballo-Rubio:2022kad}~\footnote{Rotating counterparts of spherically symmetric inner-extremal regular black holes have been constructed in~\cite{Franzin:2022wai}, cf.~e.g.~\cite{Ghosh:2022gka} for a discussion of possible observational signatures.
An early discussion of the idea of inner extremality as a possibility of avoiding mass inflation can be found in~\cite{Colleaux:2019ckh}.
}
 for generic values of the mass $M$ as the only dimensionful scale. The former types of black hole geometries are fully extremal~\cite{DiFilippo:2024spj}, in the sense that they to not contain any non-extremal horizons, whereas in the latter class of geometries the outer horizon is assumed to be non-extremal. Full extremality may be crucial for avoiding not only classical mass inflation~\cite{Carballo-Rubio:2018pmi,Bertipagani:2020awe,Carballo-Rubio:2021bpr,DiFilippo:2022qkl,Carballo-Rubio:2024dca}, but also semi-classical instabilities reflected in divergences of the renormalised stress-energy tensor in static and dynamical regular  black hole geometries~\cite{Barcelo:2020mjw,Barcelo:2022gii,McMaken:2023uue,Carballo-Rubio:2026gwg}, as well as instabilities due to Hawking evaporation~\cite{Hollands:2019whz,Balbinot:2023vcm,McMaken:2023tft,McMaken:2023uue,McMaken:2024tpc}. 
On the other hand, such geometries are expected to suffer from the classical Aretakis instability~\cite{Aretakis:2012ei}, unless the extremal horizon is of infinitely high degeneracy~\cite{Agrawal:2026oka}. The question about the eternal endpoint of a non-singular gravitational collapse is thus at present either unsettled or deceptive, if the physical evolution of a dynamical regular black hole geometry does not actually produce Cauchy and event horizons. 
We will on these grounds investigate here different scenarios of asymptotic formation of extremal regular black holes as large-mass limits of an initially horizonless spacetime. These spacetimes can also be reversely modeled to evaporate, and hence describe transient asymptotically extremal regular black holes.\\

To describe the dynamical gravitational collapse into extremal regular black holes, we will consider the coupling of the spherically reduced theory --- in all cases by definition an integrable two-dimensional Horndeski theory --- to a classical Vaidya source~\cite{Vaidya:1951zza,Vaidya:1966zza} following~\cite{Boyanov:2025pes}.~\footnote{Other models of formation of non-extremal regular black holes in polynomial and non-polynomial curvature quasi-topological gravities have been analysed e.g.~in~\cite{Bueno:2024eig,Bueno:2024zsx,Bueno:2025gjg,Bueno:2025zaj}.} The effect of such a coupling is to achieve a time dependence of the mass $M$ while remaining onshell on the equations of motion of  an associated general quasi-topological gravity. The asymptotic formation of extremal regular black can then be modeled by deforming the spacetime away from extremality together with a concrete choice of mass function which controls the timescale of the collapse. 

A single-horizon extremal regular black hole can be approached  from a horizonless object without producing a non-extremal regular black hole, as discussed kinematically e.g.~in~\cite{Carballo-Rubio:2022nuj}. For double-horizon inner-extremal regular black holes such an interpolation is less straightforward.  One may either consider the collapse into a non-extremal regular black hole and require the inner horizon(s) to become extremal sufficiently fast, or alternatively one may first produce a multi-degenerate single-horizon extremal regular black hole and subsequently envisage the formation of an inner horizon while maintaining inner-extremality at all times. Such dynamical inner-extremal regular black holes may be classically stable and semi-classically metastable~\cite{Carballo-Rubio:2026gwg}. We will comment later on the possibilities of realising these different scenarios within the dynamical framework considered here.\\

The remainder of this article is structured as follows. In Sec.~\ref{Sec:StaticExtremalRBHs} we adopt a single-function static spherically symmetric ansatz and construct asymptotically flat de Sitter core extremal regular black holes for all values of the mass $M$ as the only dimensionful scale. Concretely, we consider single-horizon extremal regular black holes in Sec.~\ref{SecSub:SingleHorizon}, and double-horizon inner-extremal regular black holes in Sec.~\ref{SecSub:DoubleHorizon}. In Sec.~\ref{Sec:QTG} we discuss how such spacetimes can interpreted as vacuum solutions to general quasi-topological gravities, and can be generalised to include a time dependence by coupling the equations of motion to a Vaidya source. Sec.~\ref{SecSub:DefQTG} reviews the implicit definition of a general quasi-topological gravity, whereas in Sec.~\ref{SecSub:VaidyaSolutions} we consider Vaidya solutions of these theories. These are compared to generalised Vaidya solutions of general relativity in Sec.~\ref{SecSub:GeneralisedVaidya}. In Sec.~\ref{Sec:SingularityTheorems} we discuss the application of the Penrose-Hawking singularity theorems in general quasi-topological gravities beyond general relativity. Sec.~\ref{SecSub:PenroseHawking} reviews the assumptions of the relevant theorems, whereas Sec.~\ref{SecSub:GeometricvsEnergyConditions} illustrates how energy conditions for generalised Vaidya spacetimes in general relativity are translated, via the kinematic convergence conditions, into dynamical onshell conditions on the theory-dependent functions characterising the spherical reduction of a given general quasi-topological gravity. Sec.~\ref{Sec:GravitationalCollapse} considers the dynamical asymptotic Vaidya collapse into a single-horizon extremal and a double-horizon inner-extremal regular black hole in Sec.~\ref{SecSub:SingleHorizonFormation} and Sec.~\ref{SecSub:DoubleHorizonFormation}, respectively. We finish with a discussion in Sec.~\ref{Sec:Conclusion}.

\section{Static extremal regular black holes}\label{Sec:StaticExtremalRBHs}

\subsection{Single-function static spherically symmetric ansatz}\label{SecSub:SSSMetrics}

Consider a four-dimensional static spherically symmetric spacetime with line element in advanced coordinates $\qty{v,r,\theta,\varphi}$ depending on a single function $f$, 
\ba\label{eq:Metricf}
g_{\mu\nu}(x)\dd{x}^\mu \dd{x}^\nu &=& -f(r) \dd{v}^2 + 2 \dd{v} \dd{r} + r^2 \dd{\Omega^2}\,.
\ea
The spacetime is assumed to be asymptotically flat and approximated by the Schwarzschild spacetime at large $r$, i.e.,
\ba\label{eq:fLarger}
f(r) &=& 1 - \frac{r_g}{r} + \mathcal{O}\qty(r^{-2})\,, \quad\quad r_g \,\,=\,\, 2M\,,
\ea
where $M$ is a positive parameter proportional to the ADM mass and geometric units are implied such that $ G_{\text{N}} = c = 1$. Regularity of polynomial curvature invariants at $r=0$ requires the metric function to satisfy $f(0)=1$ and $f'(0)=0$.
 We will concretely assume that $f$ expands at the origin as
\ba\label{eq:fSmallr}
f(r) &=& 1 - \frac{ r^2}{\ell^2 } + \mathcal{O}\qty(r^3)\,,
\ea
where $\ell $ is an effective regularisation length parameter such that the spacetime has a de Sitter core.  For such a metric to describe a regular black hole, $f$ must have an even number of real positive roots, whereby degenerate roots are counted separately. If all real positive roots of $f$ are degenerate, the geometry describes a single-horizon extremal regular black hole. We will denote the location of such a horizon by $r_*$, i.e., $r_*$ is a root of $f$ with even multiplicity $m=2n$ for $n \in \mathbb{N}$,
\ba
f(r_*) \,\,=\,\, f'(r_*) \,\, =\,\,\cdots \,\, =\,\, f^{(m-1)}(r_*) \,\,=\,\,0\,, \quad \quad f^{(m)}(r_*) \,\,\neq \,\, 0\,.
\ea
 Such a notion of extremality is distinct from the one of double-horizon inner-extremal regular black holes with a de Sitter core~\cite{Carballo-Rubio:2022kad}. The latter have an outer horizon $r_+$ corresponding to a simple root of $f$, and a separate  inner extremal horizon $r_-$ corresponding to a root of $f$ with at least cubic odd multiplicity $m = 2n+1$. We will consider examples for both types of extremal regular black holes in the next subsections.\\

A primary  motivation for considering extremal regular black holes stems from the phenomenon of mass inflation~\cite{Poisson:1989zz,Poisson:1990eh,Ori:1991zz} as a classical instability of static black holes with an inner horizon $r_-$ exhibiting non-vanishing  surface gravity
 \ba
\kappa_- &=& \frac{1}{2}f'(r_-)\,.
 \ea
Mass inflation is manifested  in an exponential growth of gravitational energy as a response to ingoing and outgoing perturbations of the geometry on a timescale $\sim \frac{1}{\abs{\kappa_-}}$, and is
generally expected to render static or slowly evolving regular black holes with a non-extremal inner horizon classically instable~\cite{Brown:2011tv,Carballo-Rubio:2018pmi,Bertipagani:2020awe,Carballo-Rubio:2021bpr,Carballo-Rubio:2022pzu,Carballo-Rubio:2024dca}.
However, if $r_-$ is a root of $f$ with multiplicity larger than one, such as in double-horizon inner-extremal regular black holes or single-horizon extremal regular black holes, then mass inflation may be avoided~\cite{Carballo-Rubio:2022kad}. We will take this as a motivation to first construct geometric models of static spherically symmetric single-horizon extremal and double-horizon inner-extremal regular black holes for generic black hole mass $M$ as the only dimensionful scale. This will in turn allow us to interpret such spacetimes as non-fine-tuned vacuum solutions of modified gravitational theories~\cite{Borissova:2026wmn,Borissova:2026krh} discussed in Sec.~\ref{Sec:QTG}, and to
model the asymptotic Vaidya collapse into extremal regular black holes by considering deformations of such theories following~\cite{Boyanov:2025pes}. 
\\

As a concrete ansatz for the construction of extremal regular black holes, we will consider a class of geometries for which $f$ is a rational function~\cite{Frolov:2016pav,Carballo-Rubio:2022kad,DiFilippo:2024spj},
\ba\label{eq:fk}
f_k(r) &=& \frac{P_k(r)}{Q_k(r)}\,,
\ea
where $P_k$ and $Q_k$ are polynomials of degree $k>2$,
\ba
P_k(r) &=&  p_k r^k + p_{k-1}r^{k-1} + \dots + p_1 r + p_0 \,,\label{eq:Pk}\\
Q_k(r) &=&  r^k + q_{k-1}r^{k-1} + \dots + q_1 r + q_0\label{Qk} \,.
\ea
In practice we will focus on the two simplest cases $k=3$ and $k=4$.
The necessary and sufficient conditions for curvature scalar regularity, $f_k(0) = 1$ and $f_k'(0)=0$, imply $p_0 = q_0$ and $p_1 = q_1$. Imposing further a de Sitter core~\eqref{eq:fSmallr}, and hence $f''(0)=- \frac{2}{\ell^2}$, fixes $p_2 = q_2 - \frac{q_0}{\ell^2}$.
The asymptotic condition~\eqref{eq:fLarger} implies $p_k = 1$, and consequently $p_{k-1}  =q_{k-1} - r_g $. The $k$ roots of $f_k$ are then determined by the roots of the polynomial
\ba
P_k(r) &=& r^k + \qty[q_{k-1}-r_g]r^{k-1} + p_{k-2}r^{k-2} + \dots  + \qty[q_2 - \frac{q_0}{\ell^2}]r^2+ q_1 r +q_0\,.
\ea
The construction of extremal regular black holes for generic black hole mass $M$ as the only dimensionful scale will in general result in a characteristic mass dependence $p_{k-i} = p_{k-i}\qty(M^i)$ and $q_{k-i} = q_{k-i}\qty(M^i)$  for $i =1,\dots,k$, while the effective regularisation length parameter enters as a function $\ell\qty(M)$. 
For future reference we note that, given such an extremal regular black hole, one may deparametrise the spacetime by reintroducing a dimensionful parameter $\ell$ as a separate scale in addition to $M$ through the identification of the coefficients $p_2\qty(M^{k-2})$, $q_2\qty(M^{k-2})$ and $q_0\qty(M^k)$ in the extremal metric, and the replacement
\ba\label{eq:Deparametrisation}
p_2\qty(M^{k-2}) & \mapsto & q_2\qty(M^{k-2}) - \frac{q_0\qty(M^k)}{\ell^2}\,.
\ea
This will allow us to connect extremal regular black holes with non-extremal ones and with horizonless compact objects. 

\subsection{Single-horizon extremal regular black holes}\label{SecSub:SingleHorizon}

\begin{figure}[t]
	\centering
	\includegraphics[width=0.7\textwidth]{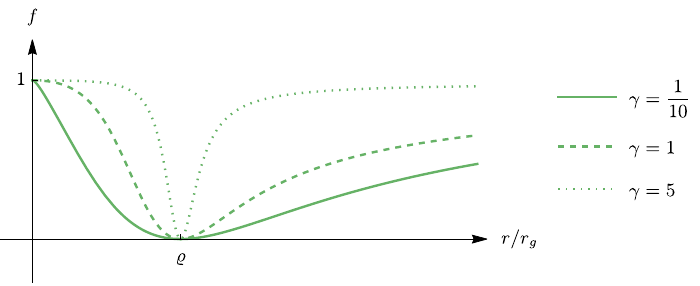}
	\caption{\label{Fig:f3Extremal} Metric function~\eqref{eq:f3Extremal} of a three-parameter family of single-horizon extremal regular black holes for fixed $\varrho = \frac{2}{3}$ and different $\gamma$. 
	}
\end{figure}

Setting $k=3$ in~\eqref{eq:fk}, the most general form of the metric compatible with the asymptotic conditions~\eqref{eq:fLarger} and~\eqref{eq:fSmallr} is~\cite{Frolov:2016pav}~\footnote{For $k=3$ the previously stated conditions imply $q_0 = r_g \ell^2$.}
\ba\label{eq:f3}
f_3(r) &=& \frac{r^3 + \qty[q_2-r_g]r^2 + q_1 r + q_0 }{ r^3 + q_2 r^2 + q_1 r + q_0}\,\, =\,\, 1 - \frac{ r_g r^2}{r^3 + q_2 r^2 + q_1 r + r_g \ell ^2}\,.
\ea
This function has three roots determined by the roots $r_\pm$ and $r_0$ of the polynomial
\ba
P_3(r) &=& r^3 + \qty[q_2 - r_g]r^2 + q_1 r + r_g \ell^2\,.
\ea
The Vieta formulae relate these to the polynomial coefficients by
\ba
r_- + r_+ + r_0 &=& - \qty[q_2 - r_g]\,,\\
\qty[r_- r_+ + r_- r_0] + \qty[r_+ r_0 ] &=& q_1\,,\\
r_- r_+ r_0 &=& 
%\qty(-1)^3
- r_g \ell^2\,.
\ea
In particular $f_3$ can have only two real positive roots $r_\pm$, and in this case the remaining one $r_0$ will be negative.
When the extremality condition $r_\pm = r_* $ is imposed, $r_*$ will be a double root in terms of which the location of the remaining simple root is given by $r_0 = - \frac{r_g \ell^2}{r_*^2}$. Then the first two equations of the above system can be solved for $q_1$ and $q_2$ as functions of $r_*$, $r_g$ and $\ell$. The result is
\ba\label{eq:qiSolution}
 \frac{q_1}{r_g^2} \,\,=\,\, \varrho^2 - 2 \frac{\gamma^2}{\varrho}\,,\quad \quad 
 \frac{q_2}{r_g} \,\,=\,\,  1 - 2 \varrho+ \frac{\gamma^2}{\varrho^2}\,,
\ea
where $\varrho = \frac{r_*}{r_g} $ and $\gamma = \frac{\ell}{r_g}$. Inserting this solution into~\eqref{eq:f3Extremal}, we have to ensure that the denominator is always positive to avoid a pole of the metric function at finite $r$. This will be the case, since the relevant polynomial can be written as 
\ba
Q_3(r) &=& r^3 +q_2 r^2 +q_1  r+ r_g \ell^2 \,\,=\,\, \frac{r_g^3}{\varrho^2}\qty[\qty(\varrho^2 \frac{r}{r_g} + \gamma^2)\qty[\frac{r}{r_g}-\varrho]^2 + \varrho^2 \qty(\frac{r}{r_g})^2]\,.
\ea
From now we will forget about their previous definition and consider the parameters $\varrho$ and $\gamma$ to be arbitrary real positive numbers.
Thereby we obtain a  three-parameter family of metrics labeled by $\varrho,\gamma \in \mathbb{R}^+$ and the mass $ M$,
\ba\label{eq:f3Extremal}
f(r) &=& 1 - \frac{2M  r^2}{r^3  + 2 \qty[1- 2 \varrho + \frac{\gamma^2}{\varrho^2}] M r^2  +4  \qty[\varrho^{2}- 2 \frac{\gamma^2}{\varrho} ] M^2  r + 8 \gamma^2  M^3 }\,,
\ea
which represent single-horizon extremal regular black holes for generic positive values of $M$. Their expansion at $r=0$ is given by
\ba\label{eq:f3ExtremalExpansion}
f(r) &=& 1 - \frac{1}{4 \gamma^2} \frac{r^2}{M^2}+ \mathcal{O}\qty(r^3)\,,
\ea
i.e., $M$ itself acts as the effective regularisation length parameter at the core. In particular this is the only dimensionful scale that these metrics contain. 
Fig.~\ref{Fig:f3Extremal} shows the metric function~\eqref{eq:f3Extremal} for fixed $\varrho $ and different $\gamma$.\\

Given a single-horizon extremal regular black hole metric~\eqref{eq:f3Extremal}, one may reversely deparametrise the spacetime to generate non-extremal regular black holes and horizonless compact objects. This can be done, for instance, by introducing a separate dimensionful regularisation parameter $\ell $ in addition to $M$ as previously described. Thus, we consider the metric function of the extremal spacetime written in the form~\eqref{eq:fk}. Therefrom we identify the coefficients $p_2\qty(M)$, $q_2(M)$ and $q_0\qty(M^3)$, and replace the former according to the prescription~\eqref{eq:Deparametrisation}. 
The case $k=3$ is special due to the constraint $p_2 = q_2 - r_g $ which follows from the requirement of a Schwarzschild limit, i.e., in effect it holds $q_0(M) = r_g \ell^2(M)$ as an identity in the extremal metric. Thus, the deparametrisation prescription can be perfomed consistently while preserving the asymptotics
through the replacement
$q_0(M) =
8 \gamma^2 M^3 \mapsto 
r_g \ell^2$ in the denominator of $f$ in~\eqref{eq:f3Extremal}.
\\

\begin{figure}[t]
	\centering
	\includegraphics[width=0.55\textwidth]{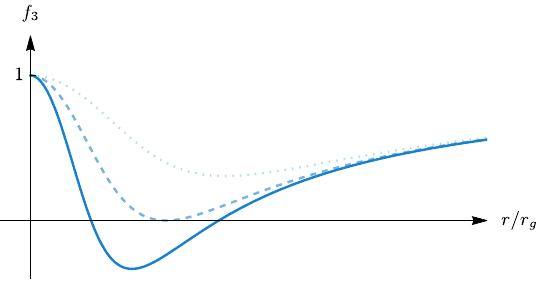}
	\caption{\label{Fig:Hayward} Metric function~\eqref{eq:f3Hayward} for different $\frac{\ell}{r_g}$, describing a non-extremal regular black hole for $\frac{\ell}{r_g} <\gamma$, an extremal regular black hole for $\frac{\ell}{r_g} =\gamma$,  and a horizonless object for $\frac{\ell}{r_g}> \gamma$.}
\end{figure}

To illustrate how the above discussion relates to standard non-extremal regular black holes with a de Sitter core and two distinct horizons $r_\pm$, we can consider the case $q_1=q_2=0$ in~\eqref{eq:f3} corresponding to the Hayward metric~\cite{Hayward:2005gi},
\ba\label{eq:f3Hayward}
f_3(r) &=&  1 - \frac{ r_g r^2}{r^3 +r_g \ell ^2}\,.
\ea
This metric
describes a double-horizon non-extremal regular black hole for $\frac{\ell}{r_g} < \gamma$, a single-horizon extremal regular black hole for $\frac{\ell}{r_g} = \gamma$, and a horizonless object for $\frac{\ell}{r_g} > \gamma$ --- where $\gamma =  \frac{2}{3 \sqrt{3}}$, cf.~Fig.~\ref{Fig:Hayward} for an illustration. The dimensionless quantity $ \frac{\ell}{r_g}$ in this context provides continuous deformation parameter on the space of Hayward metrics allowing for the interpolation between these physically distinct types of geometries.
The extremal configuration corresponds to an isolated point in this parameter space, as it is realised only for a special value of the mass $M$ compared to the regularisation length parameter $\ell$ --- namely $\ell(M) = \frac{4M}{3 \sqrt{3}}$. 
By contrast, here we are interested in geometries representing extremal regular black holes for all values of  $M$ without fine-tuning between the mass and other parameters of the model. This can be achieved by disregarding $\ell$ as a separate dimensionful parameter in addition to $M$, and instead replacing  $\ell \mapsto \ell(M)$ in~\eqref{eq:f3Hayward} to obtain
\ba\label{eq:f3HaywardExtremal}
f(r) &=& 1 - \frac{2M r^2}{r^3 + \frac{32}{27}  M^3}\,.
\ea
This metric
corresponds to the configuration labeled by $\varrho=\frac{2}{3}$ and $\gamma =\frac{2}{3 \sqrt{3}}$ of the three-parameter family of single-horizon extremal regular black holes~\eqref{eq:f3Extremal}.
 From~\eqref{eq:f3HaywardExtremal} one may also retrieve the original Hayward metric~\eqref{eq:f3Hayward} as described in the previous paragraph.~\footnote{The metric~\eqref{eq:f3HaywardExtremal} has been discussed as a solution of general relativity coupled to vector fields~\cite{Eichhorn:2025pgy}.}

\subsection{Double-horizon inner-extremal regular black holes}\label{SecSub:DoubleHorizon}

\begin{figure}[t]
 	\centering
 	\includegraphics[width=0.49\textwidth]{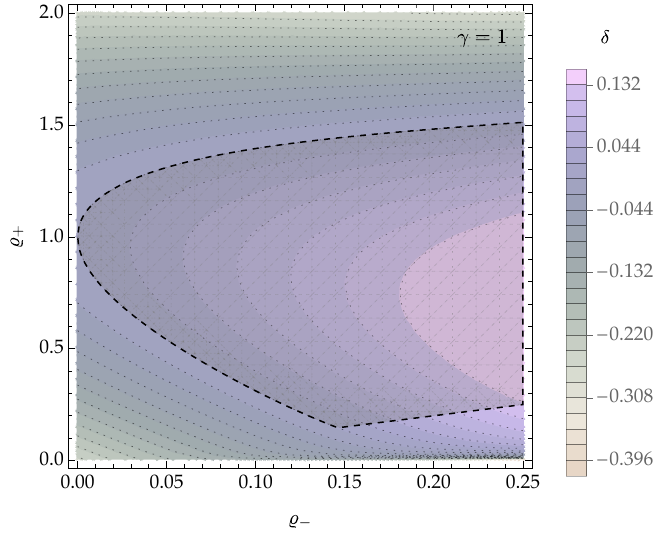}
 	\includegraphics[width=0.48\textwidth]{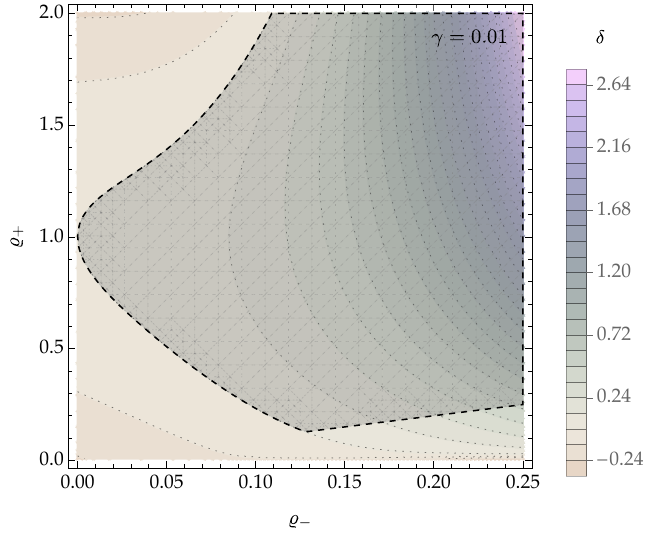}
 	\caption{\label{Fig:Delta} Parameter $\delta$~\eqref{eq:Delta} as a function of $\varrho_\pm $ for different $\gamma $. In the shaded region $\delta >0$, while also the physical requirement $\varrho_- < \varrho_+$ is satisfied. 
 	}
 \end{figure}

Setting $k=4$ in~\eqref{eq:fk}, the most general form of the metric compatible with the asymptotic conditions~\eqref{eq:fLarger} and~\eqref{eq:fSmallr} is 
\ba\label{eq:f4}
f_4(r) &=& \frac{r^4 + \qty[q_3-r_g]r^3 + \qty[q_2 - \frac{q_0}{\ell^2}]r^2+ q_1 r + q_0}{r^4 + q_3 r^3 + q_2 r^2 + q_1 r + q_0}\,.
\ea
This function has four roots determined by the roots $r_\pm$, $r_{0}$ and $r_0'$ of the polynomial
\ba
P_4(r)&=& r^4 + \qty[q_3-r_g]r^3 + \qty[q_2 - \frac{q_0}{\ell^2}]r^2+ q_1 r + q_0\,.
\ea
 The Vieta formulae relate these to the polynomial coefficients by
 \ba
 r_- + r_+ + r_0 + r_0' &=& - \qty[q_3-r_g]\,,\\
 \qty[r_- r_+ + r_- r_0 + r_- r_0'] + \qty[r_+ r_0 + r_+ r_0'] + \qty[r_0 r_0'] &=& q_2 - \frac{q_0}{\ell^2}\,,\\
 r_- r_+ r_0 + r_- r_+ r_0' + r_- r_0 r_0'+ r_+ r_0 r_0' &=& -q_1\,,\\
 r_- r_+ r_0 r_0' &=& q_0\,.
 \ea
Double-horizon inner-extremal regular black holes of the form~\eqref{eq:f4} are geometries with an outer horizon at $r_+$ corresponding to a simple root, and a separate inner extremal horizon at $r_-=r_0=r_0'$ representing a triple-degenerate root~\cite{Carballo-Rubio:2022kad}. This fixes $q_0 = r_-^3 r_+$. Then the remaining three equations of the above system can be solved for the parameters $q_1$, $q_2$ and $q_3$ as functions of $r_\pm$, $r_g$ and $\ell$. The result is
 \ba
 \frac{q_1}{r_g^3} \,\,=\,\, - \varrho_-^3  - 3 \varrho_-^2\varrho_+\,,\quad \quad 
 \frac{q_2}{r_g^2} \,\, =\,\, 3\varrho_-^2 + 3 \varrho_- \varrho_+ + \frac{\varrho_-^3\varrho_+}{\gamma^2}\,,\quad \quad 
  \frac{q_3}{r_g} \,\,= \,\,1 - 3 \varrho_- - \varrho_+\,,
 \ea
where $\varrho_\pm = \frac{r_\pm}{r_g}$ and $\gamma = \frac{\ell}{r_g}$. From now on we will forget about their definition and consider the parameters $\varrho_\pm$ and $\gamma$ to be arbitrary real positive numbers. Inserting the above solution into~\eqref{eq:f4}, we have to
ensure that the denominator is always positive to avoid a pole of the metric function at finite $r$. The relevant polynomial can be written as a sum of squares~\cite{Carballo-Rubio:2022kad} 
 \ba\label{eq:Q4}
 Q_4(r) &=&
r^4 + q_3 r^3 + q_2 r^2 + q_1 r + q_0 \nn\\
&=& \frac{1}{4}r_g^4 \qty[\qty[1+2 \frac{r}{r_g} - 3\varrho_- - \varrho_+]^2 \qty(\frac{r}{r_g} )^2 + 4 \delta \qty(\frac{r}{r_g} )^2 + \frac{\varrho_-}{\varrho_+}\qty[2 \varrho_- \varrho_+- \qty(\varrho_- + 3 \varrho_+)\frac{r}{r_g} ]^2 ]\,,\quad 
 \ea
where
$\delta$ is a dimensionless parameter  given by
 \ba\label{eq:Delta}
 \delta &=& -\frac{1}{4} \qty[3 \varrho_-^2 + 3 \varrho_-\qty[\varrho_+ -2] + \qty[1-\varrho_+]^2 + \varrho_-^3 \qty[\frac{1}{\varrho_+}- \frac{4\varrho_+}{\gamma^2}]]\,.
 \ea
For positive $\delta$ the expression in~\eqref{eq:Q4} will be positive, and thus the resulting function $f$ will have no poles on the real axis. Assuming a large separation between the two horizons such that $\epsilon = \frac{\varrho_-}{\varrho_+} \ll 1$, $\delta$ can be expanded in the form
 \ba
 \delta &=& -\frac{1}{4}\qty[\varrho_+ - 1]^2 
 - \frac{3}{4}\qty[\varrho_+ -2 ] \varrho_+ \epsilon + \mathcal{O}\qty(\epsilon^2)\,.
 \ea
 Thus, for $\varrho_+ \simeq 1$ and small but finite $\epsilon$, one may achieve $\delta >0$ for quite generic values of $\gamma $. On the other hand,  for any fixed $\varrho_\pm$, making $\gamma$ sufficiently small, the last term in~\eqref{eq:Delta} will dominate and again ensure $\delta >0$.  Fig.~\ref{Fig:Delta} shows the value of $\delta$ in the parameter space $\varrho_\pm \in \mathbb{R}^+$ at fixed $\gamma$.\\

\begin{figure}[t]
	\centering
	\includegraphics[width=0.7\textwidth]{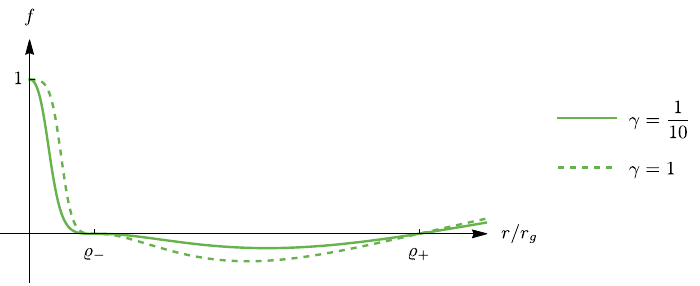}
	\caption{\label{Fig:f4Extremal} Metric function~\eqref{eq:f4Extremal} of a four-parameter family of double-horizon inner-extremal regular black holes for fixed $\varrho_- = \frac{1}{6}$ and $\varrho_+ =1$, and different $\gamma$. 
	}
\end{figure}

In summary, this discussion leads to a four-parameter family of metrics labeled by $\varrho_\pm ,\gamma \in \mathbb{R}^+$ and the mass $ M$,
\ba\label{eq:f4Extremal}
f(r) &=& \qty[r- 2 M \varrho_-]^3 \qty[r- 2 M \varrho_+] \cdot \bigg[
r^4 - 2 \qty[3 \varrho_- + \varrho_+ - 1]M r^3 + 4 \qty[3\varrho_-^2 + 3 \varrho_- \varrho_+ + \frac{\varrho_-^3 \varrho_+ }{\gamma^2}]  M^2 r^2 \quad \nn\\
& -& 8 \qty[\varrho_-^3 + 3 \varrho_-^2 \varrho_+] M^3 r + 16 \varrho_-^3 \varrho_+ M^4\bigg]^{-1}\,,
\ea
which, for suitably chosen dimensionless parameters so as to avoid a pole of $f$, represent double-horizon inner-extremal regular black holes for generic positive values of $M$. Their expansion at $r=0$  takes the form~\eqref{eq:f3ExtremalExpansion}, so that  again $M$ itself acts as the effective regularisation length parameter at the core. In particular this is the only dimensionful scale that these metrics contain.
 Fig.~\ref{Fig:f4Extremal} shows the metric function~\eqref{eq:f4Extremal} for fixed $\varrho_\pm$ and different $\gamma$.\\

\begin{figure}[t]
	\centering
	\includegraphics[width=0.55\textwidth]{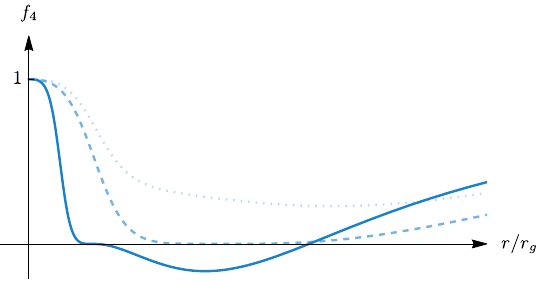}
	\caption{\label{Fig:f4NonExtremal} Interpolation from a double-horizon inner-extremal regular black hole with metric function~\eqref{eq:f4Extremal} for given $\varrho_\pm$
	%	$\varrho_- = \frac{1}{3}$, $\varrho_+ = \frac{3}{2}$ 
		and $\gamma$, to a single-horizon extremal regular black hole~\eqref{eq:f4ExtremalSingleHorizon} by taking the limit $\varrho_\pm \to \varrho \in \qty(\varrho_-,\varrho_+)$ which connects with the deparametrised metric function~\eqref{eq:f4HCOExtremal} for $\frac{\ell}{r_g} =  \gamma$, to a horizonless compact object with metric function~\eqref{eq:f4HCOExtremal} for $\frac{\ell}{r_g} > \gamma$.}
\end{figure}

Let us now consider the deformation of a double-horizon inner-extremal regular black hole~\eqref{eq:f4Extremal} into a horizonless compact object while preserving extremality.
This can be done in a two-step procedure by first deforming the spacetime into a single-horizon extremal regular black hole with a quartic root at $r_\pm = r_*$, and subsequently deparametrising the spacetime through a separate length scale $\ell$ as previously discussed. That is, from~\eqref{eq:f4Extremal} first a single-horizon extremal regular black hole is obtained by letting the horizon separation $\varrho_+ - \varrho_- \to 0$, while also requiring $\varrho_-$ to increase, which simultaneously demands $\varrho_+$ to decrease. This is  reasonable to expect in an evaporating black hole, which, conversely, during gravitational collapse would mean that after a fully extremal single-horizon regular black hole is formed, the inner horizon continues to move inwards while remaining inner-extremal at all times, whereas the outer horizon moves outwards. Denoting the quartic-degenerate root by $\varrho_\pm = \varrho$, the fully extremal metric is
\ba\label{eq:f4ExtremalSingleHorizon}
f(r) &=&  \frac{\qty[r - 2 M \varrho]^4}{r^4 + 2 \qty[1-4 \varrho]M r^3 + 4 \qty[6 \varrho^2  + \frac{ \varrho^4}{\gamma^2}]  M^2 r^2 - 32 \varrho^3 M^3 r + 16 \varrho^4 M^4}\,.\,\,\,\,\quad
\ea
In the second step, a horizonless spacetime can be obtained by following the deparametrisation prescription~\eqref{eq:Deparametrisation}, which leads to
\ba\label{eq:f4HCOExtremal}
f_4(r) &=& \frac{ \qty[r-2 M \varrho ]^4  +4  \qty[\frac{1}{\gamma^2}  - \frac{4 M^2}{\ell^2} ] \varrho^4 M^2 r^2}{r^4 + 2 \qty[1-4\varrho]M r^3  +4 \qty[6  \varrho^2  +\frac{\varrho^4 }{\gamma^2} ]M^2 r^2 - 32  \varrho^3 M^3 r + 16 \varrho^4 M^4}\,.
\ea
The single-horizon extremal regular black hole spacetime is obtained for $\ell(M) = 2 M \gamma$. Fig.~\ref{Fig:f4NonExtremal} shows this two-step interpolation.

\section{Extremal regular black holes from modified gravity}\label{Sec:QTG}

 In this section we will illustrate how the static extremal regular black holes with mass parameter $M$ considered  in the previous section can be interpreted as non-fine-tuned vacuum solutions of modified gravitational theories in four dimensions~\cite{Borissova:2026krh,Borissova:2026wmn} --- here referred to as {\it general quasi-topological gravities} --- 
%--- referred to as {\it general quasi-topological gravities}~\cite{Borissova:2026krh,Borissova:2026wmn} --- 
and can be made dynamical with a time-dependent mass function $M(v)$ by coupling these theories to a Vaidya source~\cite{Boyanov:2025pes}.

\subsection{Four-dimensional general quasi-topological gravities}\label{SecSub:DefQTG}

We will consider four-dimensional generally covariant gravitational theories
\ba\label{eq:S}
S[g] &=&  \frac{1}{16 \pi }\int \dd[4]{{x}} \sqrt{-g}\, \mathcal{L}\,,
\ea
and their reduction on spherically symmetric backgrounds
\ba\label{eq:MetricWarped}
g_{\mu\nu}(x)\dd{x}^\mu \dd{x}^\nu & =& q_{ab}(y)\dd{y}^a \dd{y}^b + \varphi(y)^2 \dd{\Omega^2},\,\,\,\quad \quad 
\dd{\Omega^2} \,\, =\,\, \gamma_{ij}(\theta)\dd{\theta}^i \dd{\theta}^j\,,
\ea
where $q_{ab}(y)$ is a two-dimensional Lorentzian metric parametrised in terms of coordinates $\qty{y^a}_{a=0,1}$, $\varphi(y) > 0$ is a scalar field, and $\gamma_{ij}(\theta) = \text{diag}\qty{1,\sin^2\theta}$ denotes the Euclidean metric on the unit two-sphere
parametrised in terms of angular coordinates $\qty{\theta^i}_{i=1,2}=\qty{\theta,\phi}$. The reduced action is given by
 \ba\label{eq:S2D}
S_{\text{2D}}[q,\varphi] &=&  \frac{1}{2} \int \dd[2]{y} \sqrt{-q} \,\mathcal{L}_{q,\varphi} \,, \quad \quad \mathcal{L}_{q,\varphi} \,\,=\,\, \frac{1}{2} \,\varphi^{d-2} \eval{\mathcal{L}}_{\eqref{eq:MetricWarped}}\,.
\ea
Four-dimensional general quasi-topological gravities~\cite{Borissova:2026krh,Borissova:2026wmn}
are covariant theories~\eqref{eq:S} whose reduced action~\eqref{eq:S2D} is a two-dimensional Horndeski action, i.e., the Lagrangian density takes the form
\ba\label{eq:S2DHorndeski}
\mathcal{L}_{q,\varphi }& =&  h_2(\varphi,\chi) - h_3(\varphi,\chi)\Box \varphi + h_4(\varphi,\chi) \mathcal{R}+ 2 \partial_\chi h_4(\varphi,\chi) \qty[\qty(\Box \varphi)^2 - \nabla_a \nabla_b \varphi \nabla^a \nabla^b \varphi]\,,
\ea
where $\chi = \nabla_a \varphi \nabla^a \varphi$ is the kinetic term of the scalar field,
and the functions $h_i$ satisfy in addition the integrability condition
\ba \label{eq:Integrability}
\partial_\chi h_2 + \partial_\varphi h_3 - \partial_\varphi^2 h_4 &=& 0
\,\,\, \quad \,\,\, \Leftrightarrow \,\,\, \quad \,\,\, \partial_\chi \alpha - \partial_\varphi \beta \,\,=\,\,0\,,
\ea
with  two functions $\alpha$ and $\beta$ defined by
\ba
\alpha(\varphi,\chi) \,\,= \,\,  h_2 + \chi \partial_\varphi \qty( h_3  - 2 \partial_\varphi h_4)\,, \,\,\, \quad \quad 
\beta(\varphi,\chi) \,\,=\,\, \chi \partial_\chi \qty(h_3  - 2 \partial_\varphi h_4) - \partial_\varphi h_4 \,.
\label{eq:AlphaBeta}
\ea
The solution to~\eqref{eq:Integrability} can be stated in terms of a generating function $\Omega$ satisfying~\cite{Boyanov:2025pes,Carballo-Rubio:2025ntd}
\ba\label{eq:OmegaDef}
\alpha(\varphi,\chi) \,\,=\,\, \partial_\varphi  \Omega(\varphi,\chi) \,, \quad \quad \beta(\varphi,\chi) \,\,=\,\, \partial_\chi \Omega(\varphi,\chi)\,,
\ea
which fully characterises spherically symmetric vacuum solutions of these gravitational theories~\cite{Borissova:2026krh,Borissova:2026wmn}. We will review the relevant results and their generalisation to dynamical Vaidya solutions which follows from~\cite{Boyanov:2025pes} in the next subsection.\\

The above definition identifies general quasi-topological gravities implicitly based on their spherical reduction.
Applied in $d\geq 5$  dimensions, this notion of quasi-topological gravity
includes polynomial curvature quasi-topological gravities~\cite{Oliva:2010eb,Myers:2010jv,Dehghani:2011vu,Cisterna:2017umf,Bueno:2019ltp,Bueno:2019ycr, Bueno:2022res,Moreno:2023rfl} as a subclass of general quasi-topological theories for which the generating function takes the form $\Omega(\varphi,\chi) = \varphi^{d-1} h(\psi)$, where $h$ is a power series in the Riemann scalar $\psi(\varphi,\chi) = (1-\chi)/\varphi^2$~\cite{Bueno:2024zsx,Bueno:2024eig,Bueno:2025qjk,Bueno:2025gjg}. The connections between polynomial curvature and more general quasi-topological gravities involving arbitrary functions of curvature and curvature-derivative invariants are established at the level of their spherical reduction in~\cite{Borissova:2026wmn,Borissova:2026krh}. The gravitational Lagrangians for higher-dimensional polynomial curvature quasi-topological theories are explicitly known in terms of analytic functions of quasi-topological densities constructed polynomially from the Riemann tensor~\cite{Bueno:2019ltp,Bueno:2019ycr,Bueno:2022res}. Such densities beyond the Einstein-Gauss-Bonnet combination do not exist in four dimensions~\cite{Bueno:2019ltp,Moreno:2023rfl}, but one may construct four-dimensional non-polynomial curvature and curvature-derivative actions giving rise to the subclasses of integrable two-dimensional Horndeski theories specified above~\cite{Bueno:2025zaj,Borissova:2026wmn,Borissova:2026krh,Colleaux:2017ibe,Colleaux:2019ckh,Colleaux:2026qew}. In particular, generic such integrable two-dimensional Horndeski theories can be reached from the spherical reduction of four-dimensional gravities~\cite{Borissova:2026wmn,Borissova:2026krh,Colleaux:2017ibe,Colleaux:2019ckh}. These are the theories to which we refer here implicitly as four-dimensional general quasi-topological gravities. The generating function $\Omega(\varphi,\chi)$ characterising the spherically reduced theory in this case can have an arbitrary dependence on the variables $\varphi$ and $\chi$~\cite{Borissova:2026krh}. The accordingly extended solution space provides the main input required for the interpretation of the extremal regular black holes constructed in Sec.~\ref{Sec:StaticExtremalRBHs} as non-fine-tuned vacuum solutions for generic $M$.\\

The variation the action~\eqref{eq:S} evaluated on~\eqref{eq:MetricWarped} takes the form 
\ba
\eval{\mathcal{E}_{\mu\nu}}_{\eqref{eq:MetricWarped}} 
&\vcentcolon = & \eval{\frac{16 \pi}{\sqrt{-g} } \frac{\var S}{\var g^{\mu\nu}}}_{\eqref{eq:MetricWarped}} \,\,=\,\,
\frac{2}{\varphi^2}\mathcal{E}_{ab} \delta^{a}_\mu \delta^b_\nu 
- \frac{\varphi}{2
} \mathcal{E}_\varphi \gamma_{ij} \delta^i_\mu \delta^j_\nu\,,
\label{eq:Emunu}\quad \,\,\,
\ea
where $\mathcal{E}_{ab}$ and $\mathcal{E}_\varphi$ are defined as  the variations of the reduced action~\eqref{eq:S2D} with respect to $q_{ab}$ and $\varphi$.
When the reduced action is a Horndeski action~\eqref{eq:S2DHorndeski} satisfying the integrability condition~\eqref{eq:Integrability}, i.e., for a general quasi-topological gravity on backgrounds~\eqref{eq:MetricWarped},  these variations are explicitly given by~\cite{Boyanov:2025pes,Carballo-Rubio:2025ntd}
\ba
\mathcal{E}_{ab} &  = &\frac{2}{\sqrt{-q} }\frac{\var S_{\text{2D}}}{\var q^{ab}} \,\, = \,\, - \frac{1}{2}\big[\alpha + 2\beta \Box \varphi\big]q_{ab}    +  \beta \nabla_a \nabla_b \varphi \,,\label{eq:EOMqab}\\
\mathcal{E}_\varphi & = &  \frac{2}{\sqrt{-q} } \frac{\var S_{\text{2D}}}{\var \varphi } \,\, = \,\, -\beta \mathcal{R}+2\partial_\varphi \beta \Box \varphi+\partial_\varphi \alpha +2\partial_\chi\beta \qty[\qty(\Box \varphi )^2-\nabla_a\nabla_b \varphi \nabla^a\nabla^b \varphi ]\,.\label{eq:EOMPhi}
\ea
The 
reduction on~\eqref{eq:MetricWarped} of an energy-momentum tensor $T_{\mu\nu}$ decomposes analogously as the variation of the gravitational action~\eqref{eq:Emunu}. Its base components thus determine the effective two-dimensional  source $t_{ab}$,
\ba\label{eq:t}
T_{ab}\,\,=\,\, \frac{2}{\varphi^2} t_{ab} \delta^a_\mu \delta^b _\nu \,\,\, \quad \,\,\, \Leftrightarrow \,\,\, \quad \,\,\, t_{ab} &=& \frac{\varphi^{2}}{2} T_{\mu\nu} \delta^\mu_a \delta^\nu_b\,,
\ea
on the right-hand side of the equations of motion for $q_{ab}$,
\ba
\mathcal{E}_{ab} &=&8 \pi  t_{ab}\,.\label{eq:EOMqabtab}
\ea
The angular components of the equations of motion $\mathcal{E}_{\mu\nu} = 8 \pi T_{\mu\nu}$
are redundant due to the  Bianchi identity implied by general covariance.\\

A rank-two second-order algebraically conserved tensor $\mathcal{G}_{\mu\nu}(q_{ab},\varphi)$ given by the right-hand side of~\eqref{eq:Emunu}, and therefrom effective field equations $\mathcal{G}_{\mu\nu} = 8 \pi T_{\mu\nu}$ have been constructed originally as deformations of the Einstein equations on spherically symmetric backgrounds which remain of second order in derivatives of the gravitational field and allow for effective analyses of black hole interiors beyond general relativity~\cite{Carballo-Rubio:2025ntd}. The above tensor can be generated as the variation $\mathcal{E}_{\mu\nu}$ of a generally covariant gravitational action evaluated on~\eqref{eq:MetricWarped} for generic $q_{ab}$ and $\varphi$~\cite{Borissova:2026wmn,Borissova:2026krh}.
%As such its conservation follows from general covariance. 
When restricted to the subclass of two-dimensional Horndeski theories satisfying the integrability condition~\eqref{eq:Integrability}, the deformed Einstein equations~\cite{Carballo-Rubio:2025ntd} are then in fact the spherically reduced equations of motion of general quasi-topological gravities minimally coupled to matter. This statement directly generalises to higher-dimensional polynomial curvature quasi-topological gravities taking into account the specific form of the functions $\alpha$ and $\beta$ in this case~\cite{Borissova:2026wmn,Borissova:2026krh}. In particular, solving the
effectively two-dimensional equations~\eqref{eq:EOMqabtab} amounts to finding spherically symmetric solutions to such modified gravitational theories. In practice, we will deal here with the equations~\eqref{eq:EOMqabtab} without asking about the actual form of the four-dimensional generally covariant action. A analogue viewpoint has been adopted in~\cite{Borissova:2026rbi} to derive the thermodynamics of spherically symmetric vacuum black holes in these theories.

\subsection{$2D$ Horndeski equations of motion and Vaidya solutions}\label{SecSub:VaidyaSolutions}

Let us now consider a general dynamical spherically symmetric metric in advanced coordinates $\qty{v,r,\theta,\varphi}$ parametrised by two functions $f$ and $n>0$,
\ba\label{eq:Metricnf}
\dd{s}^2 &=& -n(v,r)^2 f(v,r) \dd{v}^2  + 2 n(v,r)\dd{v}\dd{r} + r^2 \dd{\Omega^2}\,,
\ea
which amounts to setting
\ba\label{eq:qPhinf}
q_{ab}(y) \dd{y}^a \dd{y}^b\,\,=\,\,  -n(v,r)^2 f(v,r) \dd{v}^2  + 2 n(v,r)\dd{v}\dd{r} \,, \quad \quad \varphi(y) \,\,=\,\, r
\ea
in~\eqref{eq:MetricWarped}. In this case $\chi(v,r) = f(v,r)$, and the independent equations of motion~\eqref{eq:EOMqabtab} are~\cite{Borissova:2026dlz}
\ba
\frac{n}{2} \qty[n f \qty(\alpha + \beta \partial_r f) +  \beta \partial_v f] &=& 8 \pi t_{vv}\,,\\
- \frac{n}{2} \qty[\alpha + \beta \partial_r f]&=& 8 \pi t_{vr}\,,\\
 - \frac{1}{n} \beta \partial_r n  &=& 8 \pi t_{rr}\,,
\ea
where $\alpha$ and $\beta$ are given in terms of the generating function $\Omega$ defined in~\eqref{eq:OmegaDef} as
\ba\label{eq:OmegaDeff}
\alpha(r,f) &=&  \partial_r \Omega(r,f) \,, \quad \quad \beta(r,f) \,\,=\,\, \partial_f \Omega(r,f)\,,
\ea
and $\beta \neq 0$ is assumed throughout.
Following~\cite{Boyanov:2025pes}, we will now particularise to a Vaidya source~\cite{Vaidya:1951zza,Vaidya:1966zza} with energy-momentum tensor 
\ba \label{eq:TVaidya}
T_{\mu\nu} \,\,=\,\, \mu \partial_\mu v \partial_\nu v \,\,\, \quad \,\,\, \Leftrightarrow \,\,\, \quad \quad t_{ab} 
\,\,= \,\,
 \frac{r^2}{2} \mu \partial_a v \partial_b v\,,
\ea
where the energy density $\mu$ is given by
\ba
\mu(v,r) &=&  \frac{\dot{M}(v)}{4 \pi r^2}\,,
\ea 
with a time-dependent mass function $M(v)>0$.
The above set of equations then simplifies to~\cite{Borissova:2026dlz}
\ba\label{eq:EqsEvaluated}
\derivative{}{r} \Omega(r,f)\,\, = \,\, \alpha + \beta \partial_r f \,\,=\,\, 0\,, \quad \quad 
\beta n\partial_v f\,\,=\,\, 2 \dot{M} \,,\quad \quad 
\partial_r n \,\,=\,\, 0\,.
\ea
The last equation implies $n=n(v)$, and hence one may set $n=1$ by a redefinition of the coordinate $v$. A Vaidya source in particular preserves the single-function form of spherically symmetric vacuum solutions and accordingly Birkhoff theorem in general quasi-topological gravities~\cite{Borissova:2026krh,Borissova:2026wmn}, including the one applicable in higher-dimensional polynomial curvature quasi-topological gravities~\cite{Bueno:2025qjk}. See moreover~\cite{Frolov:2026tft} for a recent discussion of Vaidya solutions in this latter class of theories.\\

The general solution to the above set of equations can be found by integrating the first one to obtain $\Omega(r,f) =
\omega(v)$ for
an arbitrary function $\omega$ of $v$ to be fixed. Taking the $v$-derivative of this relation and therein inserting the second equation, it follows that $\dot{\omega} = 2 \dot{M}$. Therefore,
the following algebraic equation~\cite{Boyanov:2025pes} determines the metric function $f(r,v)$ of a Vaidya solution to a general quasi-topological gravity,
\ba\label{eq:EqAlgebraic}
\Omega(r,f) &=& 2 M(v)\,,
\ea
where the integration constant has been set to zero.
\\

In the remainder of this work we will adopt the following perspective. Given a static asymptotically flat metric~\eqref{eq:Metricf} with function $f(r)$ depending on the mass parameter $M$, this spacetime can be reconstructed as a solution to the equations of motion of an integrable two-dimensional Horndeski theory~\cite{Boyanov:2025pes,Carballo-Rubio:2025ntd}, and hence as a vacuum solution to a general quasi-topological gravity~\cite{Borissova:2026krh,Borissova:2026wmn}. The generating function $\Omega$ characterising the spherical reduction of an associated quasi-topological theory is formally determined by the inversion $M=M(r,f)$, and the identification $\Omega(\varphi,\chi) = 2 M(\varphi,\chi)$ obtained by replacing $r \mapsto \varphi$ and $f \mapsto \chi$ offshell. Then the time-dependent generalisation arising from the coupling of this  quasi-topological gravity to a Vaidya source amounts to the replacement $M \mapsto M(v)$ in $f$.
In particular, in order to make a given static spacetime~\eqref{eq:Metricf} with mass parameter $M$ dynamical with a time-dependent mass function $M(v)$ while remaining onshell on the equations of motion of such a modified gravitational theory, an inversion of $f$ for $M$ is not necessary.~\footnote{Such an inversion would be required if the two-dimensional Lagrangian is desired in explicit form~\cite{Boyanov:2025pes,Borissova:2026dlz}, so that it can subsequently be lifted to a four-dimensional general quasi-topological gravity~\cite{Borissova:2026krh,Borissova:2026wmn}. Here we do not discuss such gravitational theories explicitly, and instead remain within the effectively two-dimensional Horndeski framework reviewed in this subsection.} Thus, in summary, we will interpret static spherically symmetric spacetimes~\eqref{eq:Metricf} with mass parameter $M$ as vacuum solutions of general quasi-topological theories, and their time-dependent versions obtained by replacing $M \mapsto M(v)$ in the metric as Vaidya solutions of these same theories. This perspective is then applied to extremal regular black holes as the main focus in this work. The primary motivation for constructing extremal regular black holes for generic mass $M$ and free from additional dimensionful scales in Sec.~\ref{Sec:StaticExtremalRBHs} is then i) that a possible required fine-tuning between $M$ and potential coupling constants of the theory, which would be present in the generating function $\Omega$, is avoided, i.e., extremality of the resulting solutions is achieved as a generic feature of a given theory rather than at isolated points of the solution space, and that ii) subsequently coupling the theory to a Vaidya source makes all horizons simultaneously dynamical.

\subsection{Relation to generalised Vaidya solutions of general relativity}\label{SecSub:GeneralisedVaidya}

In this subsetion we will elaborate on the relation between vacuum and Vaidya solutions of general quasi-topological gravities discussed above based on~\cite{Boyanov:2025pes,Carballo-Rubio:2025ntd,Borissova:2026wmn,Borissova:2026krh}, and generalised Vaidya solutions of general relativity~\cite{Husain:1995bf,Wang:1998qx}. The latter are spherically symmetric spacetimes
\ba\label{eq:MetricfDynamical}
\dd{s}^2 &=& -f(v,r)\dd{v}^2 + 2 \dd{v}\dd{r} + r^2 \dd{\Omega^2}\,, \quad \quad f(v,r) \,\,=\,\, 1 - \frac{2 m(v,r)}{r}\,,
 \ea
where $m$ is generalised Misner-Sharp mass function~\cite{Misner:1964je},
 which satisfy the Einstein equations $G_{\mu\nu} = 8 \pi T_{\mu\nu}$ with an energy-momentum tensor representing a type II fluid~\cite{Hawking:1973uf},
\ba\label{eq:TGeneralisedVaidya}
T_{\mu\nu} &=& \mu l_\mu l_\nu + \qty(\rho + p)\qty[l_\mu n_\nu + l_\nu n_\mu ] + pg_{\mu\nu}\,,
\ea
where $l^\mu$ and $n^\mu$ are two base null vectors satisfying $l_\mu n^\mu = -1$.
The contributions to the energy density $\mu$ and $\rho$, and the pressure $p$ are given by
\ba\label{eq:MuRhoP}
\mu(v,r) \,\,=\,\, \frac{\dot{m}(v,r)}{4 \pi r^2}\,, \quad \quad \rho(v,r) \,\,=\,\, \frac{m'(v,r)}{4 \pi r^2}\,, \quad \quad p(v,r) \,\,=\,\, -\frac{m''(v,r)}{8 \pi r}\,.
\ea
In other words, any spacetime of the form~\eqref{eq:MetricfDynamical} can be generated as a solution to the Einstein equations with energy-momentum tensor~\eqref{eq:TGeneralisedVaidya}, and this applies in particular to  static regular black hole spacetimes obtained as vacuum solutions to general quasi-topological gravities, i.e., spacetimes satisfying~\eqref{eq:EqAlgebraic} with an integration constant $M$, or more generally Vaidya solutions to such theories with a time-dependent mass function $M(v)$. In this case, $f$ will be a function of the form $f(M(v),r)$ obtained by solving the algebraic equation~\eqref{eq:EqAlgebraic}, which in turn defines the effective generalised Misner-Sharp mass function $m=m(M(v),r)$ in~\eqref{eq:MetricfDynamical} by
\ba\label{eq:fVaidyaQTG}
f\qty(v,r) &=& 1 - \frac{2 m\qty(M(v),r)}{r}\,.
\ea
Then the matter variables~\eqref{eq:MuRhoP} take the form
\ba\label{eq:MuRhoPQTG}
\mu(v,r) \,\,=\,\,\frac{\dot{M}(v) \partial_M m }{4 \pi r^2}\,, \quad \quad  \rho(v,r) \,\,=\,\, \frac{m'(M(v),r)}{4 \pi r^2}\,, \quad \quad p(v,r) \,\,=\,\, -\frac{m''(M(v),r)}{8 \pi r}\,.
\ea
In the static case, when $\mu$ vanishes, 
$\rho$ and $p$ represent the  energy density and pressure which would be required for  sourcing a vacuum solution of a general quasi-topological gravity as a static generalised Vaidya solution of general relativity. The main point of modifying general relativity into a general quasi-topological gravity is then that the above effective energy density $\rho$ and pressure $p$ are induced by gravity itself rather than by the addition of extra matter degrees of freedom.
These effective matter variables can be expressed in terms of the gravitational functions $\alpha$ and $\beta$ characterising the spherical reduction of a given such modified theory. To that end, using~\eqref{eq:fVaidyaQTG} and~\eqref{eq:MuRhoPQTG}, as well as the corresponding equation of motion~\eqref{eq:EqsEvaluated}, one may compute
\ba
\psi -   8 \pi   \rho &=& \frac{f'}{r} \,\,=\,\,- \frac{1}{r}\frac{\alpha}{\beta} 
\,,\label{eq:fPrime}\\
 2 \psi - 16 \pi (\rho + p) &=& - f''\,\,= \,\,\frac{1}{\beta^3} \qty[\alpha^2 \partial_f \beta + \beta^2 \partial_r \alpha - 2 \alpha \beta \partial_r \beta]
 \,,\label{eq:fPrimePrime}
\ea
where we remind the reader that $\alpha = \partial_r \Omega$ and $\beta = \partial_f \Omega$, whereas $\psi(r,f) = \frac{1-f}{r^2} $.
Therefrom we obtain
\ba
8 \pi \rho \,\,=\,\, \psi + \frac{1}{r} \frac{\alpha}{\beta}\,,\quad \quad 
16 \pi p \,\,=\,\,-\frac{1}{\beta^3}\qty[ \alpha^2 \partial_f \beta  + \beta^2 \partial_r \alpha  -2 \alpha \beta \partial_r \beta  + \frac{2}{r}\alpha \beta^2   ]\,.\label{eq:RhoP}
\ea
If the generating function takes a form resembling the four-dimensional analogue of the generating function for polynomial curvature quasi-topological gravities in $d\geq 5$ dimensions, i.e.,~\footnote{This case, and accordingly the two-dimensional Horndeski theories obtained from the reduction of higher-dimensional polynomial curvature quasi-topological gravities~\cite{Bueno:2024zsx,Bueno:2025gjg,Bueno:2025qjk}, are, however, of minor relevance in this work, as the resulting generating functions $\Omega$ should not be expected to give rise to extremal black holes generically without fine-tuning --- see e.g.~\cite{DiFilippo:2024mwm} emphasising the underlying obstruction for double-horizon inner-extremal regular black holes in these theories, and~\cite{Liu:2026ltw} for an analogous discussion in the context of polymerised vacuum solutions motivated by loop quantum gravity in four dimensions. Instead, our discussion requires the extension of the two-dimensional theory space to include generic integrable two-dimensional Horndeski theories and thereby allow for a general dependence of $\Omega$ on $r$ and $f$~\cite{Borissova:2026krh}.}
\ba\label{eq:OmegaCurvature}
\Omega(r,f) &=& r^{3}h(\psi(r,f)) \,\,\, \quad \,\,\, \Rightarrow \,\,\, \quad \,\,\, \alpha(r,f) \,\,=\,\, r^2 \qty[3 h - 2 \psi h']\,,  \,\,\, \quad \,\,\,\beta(r,f) \,\,=\,\, - r h'\,,
\ea
where a prime denotes the derivative with respect to the argument $\psi$,
such as in the theories considered in~\cite{Bueno:2025zaj, Borissova:2026wmn}, the above expressions reduce to
\ba
8 \pi \rho \,\,=\,\, 3\psi - \frac{3 h}{h'}\,,\quad \quad 
16 \pi p \,\,=\,\, -6 \psi + \frac{6 h}{h'} + \frac{9 h^2 h''}{h'^3}\,.
\ea
For $h(\psi) =\psi$, corresponding to general relativity with vanishing cosmological constant, these expressions become $\rho =0$ and $p=0$ consistently with classical Vaidya solutions~\cite{Vaidya:1951zza,Vaidya:1966zza}. More generally, the above expressions provide a translation between vacuum solutions of general quasi-topological gravities, and the corresponding static generalised Vaidya solutions describing the same spacetime but sourced in the context of the Einstein equations. This provides a basis for translating energy conditions in general relativity into onshell conditions on the theory-dependent functions $\alpha$ and $\beta$ characterising a given spherically reduced general quasi-topological gravity. We will connect such a discussion with the geometric convergence conditions required in applications of the singularity theorems in the next section.

\section{Application of singularity theorems beyond general relativity}\label{Sec:SingularityTheorems}

In this section we will illustrate the correspondence between energy conditions in general relativity, and geometric convergence conditions reformulated dynamically as onshell conditions in general quasi-topological gravities. We will focus concretely on spherically symmetric vacuum and Vaidya solutions to these latter theories, and illustrate how the violation of energy conditions required for regular black holes in general relativity is triggered by the modified gravitational dynamics in general quasi-topological theories. A detailed analysis including more general matter in quasi-topological gravities is beyond the scope of this work. See, however,~\cite{Bueno:2026dln,Arrechea:2026ngi,Carballo-Rubio:2026mvj} for complementary analyses addressing conditions for curvature scalar regularity and geodesic completeness in such theories in the presence of matter.

\subsection{Review of Penrose-Hawking singularity theorems}\label{SecSub:PenroseHawking}

We will now first review the two relevant singularity theorems. The Penrose 1965 singularity theorem~\cite{Penrose:1964wq,Hawking:1973uf} establishes the null geodesic incompleteness of a spacetime under the following conditions, i) existence of a non-compact Cauchy hypersurface, ii) existence of a closed trapped surface, and iii) the null convergence condition (NCC) for all null vectors $k^\mu$,
\ba
R_{\mu\nu}k^\mu k^\nu & \geq  & 0\,.\label{eq:NCC}
\ea
When the gravitational dynamics is governed by the Einstein equations,
$G_{\mu\nu} = 8 \pi T_{\mu\nu}$,
the NCC becomes equivalent to the null energy condition (NEC) for all null vectors $k^\mu$, 
\ba
T_{\mu\nu}k^\mu k^\nu &\geq & 0\,. \label{eq:NECT}
\ea
For a generalised Vaidya energy-momentum tensor~\eqref{eq:TGeneralisedVaidya},
writing a general null vector as $k^\mu =a  l^\mu +b n^\mu + k_\perp^\mu$, where $k_\perp \cdot l = k_\perp \cdot n =0$ and $\qty(k_{\perp} )^2 = 2 a b \geq 0$, one may compute
\ba
T_{\mu\nu}k^\mu k^\nu &=&  \mu b^2 + 2 \qty(\rho + p) ab \,.
\ea
Thus the NEC becomes
\ba\label{eq:NEC}
\mu \,\, \geq \,\, 0 \,, \quad \quad \rho + p \,\,\geq \,\,0\,.
 \ea
The first condition imposes $\dot{m} \geq 0$, i.e., for instance, an apparent horizon described implicitly by the equation $r = 2 m$ must not contract with time. 
The NEC, in what concerns the application of the Penrose theorem, is only meaningful when the Einstein equations hold. More generally one should refer to the geometric NCC when applying this theorem. \\

The second relevant theorem is the Penrose-Hawking 1970 singularity theorem~\cite{Hawking:1970zqf,Hawking:1973uf}, which establishes the causal, i.e., timelike or null,
geodesic incompleteness of a spacetime  under the following conditions, i) generic curvature condition, ii) chronology condition, iii) existence of a closed trapped surface, and iv) the timelike convergence condition (TCC) for all timelike vectors $v^\mu$,
\ba
R_{\mu\nu}v^\mu v^\nu & \geq  & 0\,.\label{eq:TCC}
\ea
The TCC contains the NCC as a limiting case when $v^\mu$ becomes a null vector.
Under the assumption that the gravitational dynamics is governed by the Einstein equations, $G_{\mu\nu} = 8 \pi T_{\mu\nu}$, the trace equation can be used to express the 
Ricci tensor onshell in terms of the energy-momentum tensor and its trace. As a result, the 
TCC can be rewritten as the strong energy condition (SEC) for all timelike vectors $v^\mu$,
\ba\label{eq:SECT}
\qty[T_{\mu\nu} - \frac{1}{2} T g_{\mu\nu} ]v^\mu v^\nu &\geq & 0 \,.
\ea
For a generalised Vaidya energy-momentum tensor~\eqref{eq:TGeneralisedVaidya},
writing a general unit-normalised timelike vector as $v^\mu = a l^\mu + b n^\mu + v_\perp^\mu$, where $v_\perp \cdot l = v_\perp \cdot n = 0$ and $ \qty(v_\perp)^2= 2 ab -1 \geq 0$, one may compute
\ba
T_{\mu\nu}v^\mu v^\nu \,\,=\,\,
 \mu b^2 + \rho + (\rho +p) \qty(v_\perp)^2\,,\quad \quad 
T \,\,=\,\,
 2 \qty(p - \rho)\,,
\ea
and so
\ba
\qty[T_{\mu\nu} - \frac{1}{2} T g_{\mu\nu} ]v^\mu v^\nu 
&=&   \mu b^2 + p + (\rho +p) \qty(v_\perp)^2\,.
\ea
Thus the SEC becomes
\ba\label{eq:SEC}
\mu\,\,\geq \,\, 0\,, \quad \quad \rho + p \,\, \geq \,\, 0\,, \quad \quad  p \,\, \geq \,\, 0\,.
\ea
The first two conditions are already imposed by the NEC. For more detailed discussions of energy conditions for type II fluids, see e.g.~\cite{Hawking:1970zqf,Wang:1998qx,Maeda:2022vld}.
The SEC, in what concerns the application of the Penrose-Hawking theorem, is only meaningful when the Einstein equations hold. More generally one should refer to the geometric TCC when applying this theorem.

 \subsection{Correspondence between geometric convergence and energy conditions}\label{SecSub:GeometricvsEnergyConditions}

A detailed discussion of geometric convergence conditions for spherically symmetric regular black hole spacetimes can be found in~\cite{Borissova:2025msp,Borissova:2025hmj,Borissova:2026ohj}. For self-consistency we will rederive the relevant geometric inequalities and subsequently evaluate these for examples of extremal regular black holes discussed in Sec.~\ref{Sec:StaticExtremalRBHs}. Our main goal is to
emphasise how modifying gravity into a general quasi-topological theory can systematically trigger a dynamical violation of convergence conditions necessary for regular black holes, without requiring a breaking of the strong energy condition as is necessary in general relativity. \\

The Ricci tensor for the metrics~\eqref{eq:MetricfDynamical} has components 
\ba
R_{vv} + \frac{\dot{f}}{r}\,\,=\,\, -f R_{vr} \,\, = \,\, \frac{f}{2 r} \qty[2 f' + r f'']\,, \quad \,\,\,
R_{rr} \,\,=\,\, 0\,,\quad \,\,\,
R_{ij} \,\,=\,\, \qty[1-f - r f'] \gamma_{ij}\,.
\ea
 In spherical symmetry one  angular component of a vector can be set to zero without loss of generality. 
A general ansatz for a timelike vector is therefore
\ba
v^\mu &=& v^v \qty(\partial_v)^\mu + v^r \qty(\partial_r)^\mu + v^\theta \qty(\partial_\theta)^\mu\,.\label{eq:vfAnsatz}
\ea
Imposing the normalisation condition
\ba
g_{\mu\nu}v^\mu v^\nu  &=&  - f \qty(v^v)^2 + 2 v^v v^r + r^2 \gamma_{\theta\theta} \qty(v^\theta)^2\,\,=\,\, -\eta\,,\label{eq:vfNormalisation}
\ea
for some $\eta > 0$, 
and rescaling $ v^v \to 1$, leads to
\ba\label{eq:vf}
v^\mu &=& \qty(1, - \frac{1}{2}\qty[\eta - f  +  \qty(v^\theta)^2 r^2], v^\theta, 0)\,.
\ea
Repeating the analogous steps for a null vector $k^\mu$ by setting the right-hand side of equation~\eqref{eq:vfNormalisation} to zero, i.e., taking the limit $\eta \to 0$ of~\eqref{eq:vf}, results in
\ba\label{eq:kf}
k^\mu &=& \qty(1,  \frac{1}{2}\qty[ f -  \qty(k^\theta)^2 r^2], k^\theta, 0)\,.
\ea
This null vector interpolates between the two radial null vectors
\ba
k_-^\mu \,\,=\,\,-\qty(\partial_r)^\mu \,\,=\,\, \frac{2}{r^2}\lim_{k^\theta \to \infty} \qty[\frac{1}{\qty(k^\theta)^2} k^\mu]\,, \quad \quad 
k_+^\mu \,\,=\,\, \qty(\partial_v)^\mu + \frac{f}{2} \qty(\partial_r)^\mu \,\,=\,\, \lim_{k^\theta \to 0} k^\mu\,,
\ea
tangent to ingoing and outgoing null geodesics. The contraction of the Ricci tensor with these two null vectors is
\ba
R_{\mu\nu}k_-^\mu k_-^\nu \,\,=\,\,
0\,, \quad \quad R_{\mu\nu}k_+^\mu k_+^\nu \,\,=\,\, - \frac{\dot{f}}{r}\,,
\ea
indicating that the NCC for purely radial vectors in single-function static spacetimes~\eqref{eq:Metricf} is always marginally satisfied regardless of the specific form of the function $f$. 
This applies in particular to de Sitter core static regular black holes
and to anti-de Sitter core counterparts~\cite{Arrechea:2025nlq,Borissova:2025hmj}. Such  single-function static spherically symmetric regular black holes, in fact, do not necessarily violate the NCC, but in any case circumvent the Penrose theorem by violating the first assumption of global hyperbolicity~\cite{Borissova:2025msp}. However, any such regular black hole satisfying the weaker causality and other assumptions of the Penrose-Hawking theorem must necessarily violate the TCC~\cite{Borissova:2025hmj}. In addition, transient regular black holes generally violate the NCC in the evaporation process of the trapped region~\cite{Borissova:2026ohj}.\\

Contracting the Ricci tensor with the timelike vector~\eqref{eq:vf} results in
\ba\label{eq:Rvfvf}
R_{\mu\nu}v^\mu v^\nu &=& \qty[ \frac{f'}{r} + \frac{1}{2}f'']\eta  +  \qty[\psi  + \frac{1}{2}f'']\qty(v^\theta)^2 r^2 - \frac{\dot{f}}{r}\,.
\ea
The TCC requires semi-positivity of this expression for generic $\eta$ and $v^\theta$,  which is satisfied if  
\ba
\frac{f'}{r} + \frac{1}{2}f'' \,\, \geq \,\, 0\,, \quad \quad 
\psi + \frac{1}{2}f''  
\,\, \geq \,\, 0\,,\quad \quad 
-\frac{\dot{f}}{r} \,\, \geq \,\, 0\,.\label{eq:TCCReducedf}
\ea
The last two conditions are already imposed by the NCC, as follows by 
considering the limit $\eta \to 0$ in which $v^\mu$ becomes a null vector $k^\mu$.\\

\begin{figure}[t]
	\centering
	\includegraphics[width=0.75\textwidth]{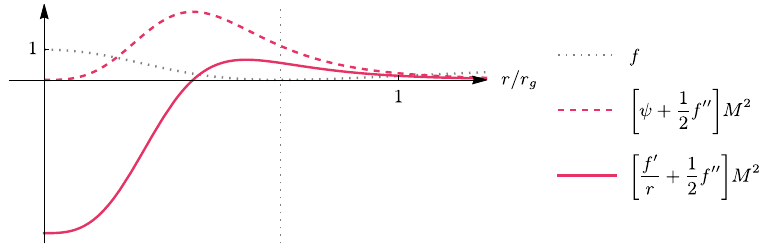}
	\caption{\label{Fig:HaywardExtremalTCC} Geometric convergence conditions~\eqref{eq:TCCReducedf} for a static single-horizon extremal regular black hole~\eqref{eq:f3HaywardExtremal}. While the NCC~\eqref{eq:NCC} is nowhere violated, the TCC~\eqref{eq:TCC} is violated locally near the core and reaches a value $- \frac{3}{\ell^2}$ at $r=0$, where $\ell$ is the effective regularisation length parameter defined by~\eqref{eq:fSmallr}, and is proportional to $\sim \frac{1}{M}$ for the extremal regular black holes considered here. The dashed line marks the location of the extremal horizon given by $\varrho = \frac{2}{3}$.}
\end{figure} 

\begin{figure}[t]
	\centering
	\includegraphics[width=0.75\textwidth]{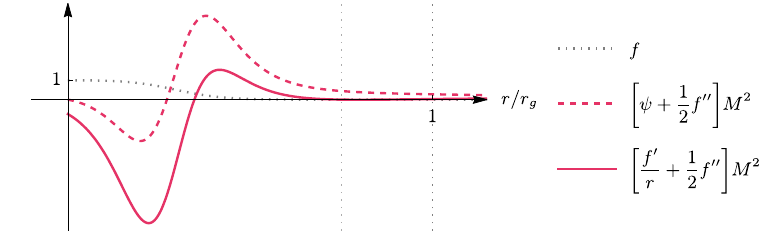}
	\caption{\label{Fig:InnerExtremalTCC} Geometric convergence conditions~\eqref{eq:TCCReducedf} for a static double-horizon inner-extremal regular black hole~\eqref{eq:f4ExtremalSingleHorizon} with free parameters set to $\varrho_- = \frac{3}{4}$, $\varrho_+=1$ and $\gamma=1$. Not only the TCC~\eqref{eq:TCC} but also the NCC~\eqref{eq:NCC} is violated locally near the core. The TCC violation reaches a value $- \frac{3}{\ell^2}$ at $r=0$, where $\ell$ is the effective regularisation length parameter defined by~\eqref{eq:fSmallr}, and is proportional to $\sim \frac{1}{M}$ for the extremal regular black holes considered here. The dashed lines mark the locations $\varrho_\pm$ of the inner extremal and outer non-extremal horizons.}
\end{figure} 

 Fig.~\ref{Fig:HaywardExtremalTCC} and Fig.~\ref{Fig:InnerExtremalTCC} show the expressions on the left-hand sides of the first two conditions above for a static single-horizon extremal regular Hayward black hole~\eqref{eq:f3HaywardExtremal}, and a static double-horizon inner-extremal regular black hole~\eqref{eq:f4Extremal} for fixed values of the free parameters.  While both geometries represent static de Sitter core regular black holes, in the former spacetime only the TCC but not the NCC is violated at and near the core, while in the latter also the NCC is violated near the core and saturated at $r=0$. 
  The TCC violation at $r=0$ in any de Sitter core regular black hole can be explained by the fact that  $f(r)=1 - \frac{r^2}{\ell^2} + \mathcal{O}\qty(r^3)$ by assumption, and
 hence $\frac{f'(r)}{r} = - \frac{2}{\ell^2} + \mathcal{O}\qty(r)$ while $f''(r) =  - \frac{2}{\ell^2} + \mathcal{O}\qty(r)$. The quantity relevant for the first inequality in~\eqref{eq:TCCReducedf} then becomes
\ba
\eval{\frac{f'}{r} + \frac{1}{2}f''}_{\eqref{eq:fSmallr}} &=& - \frac{3}{\ell^2} + \mathcal{O}(r)\,,
\ea
and hence the TCC is universally violated in any de Sitter core regular black hole. On the other hand, the violation of the NCC in the example of a static double-horizon inner-extremal regular black hole considered in Fig.~\ref{Fig:InnerExtremalTCC} takes place in a neighborhood around but away from the core, and hence requires a beyond-leading-order expansion. This NCC violation is present for any of the static double-horizon inner-extremal regular black hole models~\eqref{eq:f4Extremal}, since for small $r$,
\ba
\eval{\psi + \frac{1}{2}f''}_{\eqref{eq:f4Extremal}} &=& - \frac{\qty[\varrho_-^3 + 3 \varrho_-^2 \varrho_+ + \gamma^2]}{4 M^3 \varrho_-^3 \varrho_+ \gamma^2}r + \mathcal{O}\qty(r^2)\,.
\ea
By contrast, for the static single-horizon extremal regular black holes~\eqref{eq:f3Extremal}, a potential NCC violation near the core is model-dependent, since in this case for small $r$,
\ba
\eval{\psi + \frac{1}{2}f''}_{\eqref{eq:f3Extremal}} &=& \frac{\qty[\varrho^3- 2 \gamma^2]}{4 M^3 \varrho \gamma^4}r  + \mathcal{O}\qty(r^2)\,.
\ea
The numerical prefactor in front of the first term vanishes identically for the extremal Hayward black hole~\eqref{eq:f3HaywardExtremal}. However, other models may exhibit NCC violation near the core if this factor is negative. Both expressions above illustrate that static single-horizon extremal regular black holes with a de Sitter core may  well violate not only the TCC but also the NCC.\\

Inserting for $f'$ and $f''$ the left-hand side identities in~\eqref{eq:fPrime} and~\eqref{eq:fPrimePrime} involving $\rho$ and $p$, and making use of~\eqref{eq:MuRhoP} to express $\dot{f}$ in terms of $\mu$, one may verify that the geometric conditions~\eqref{eq:TCCReducedf} are equivalent to the conditions~\eqref{eq:SEC} imposed by the SEC on the energy-momentum tensor~\eqref{eq:TVaidya}. This must be the case, given that generalised Vaidya spacetimes are solutions of general relativity and hence the SEC is then equivalent to the TCC. On the other hand, using the right-hand side identities in~\eqref{eq:fPrime} and~\eqref{eq:fPrimePrime}, one may translate the evaluated TCC~\eqref{eq:TCCReducedf}, or equivalently the evaluated SEC in general relativity~\eqref{eq:SEC}, into onshell conditions on the theory-dependent functions $\alpha $ and $\beta $ characterising the spherical reduction of a given general quasi-topological gravity. For static spacetimes the relevant conditions are
\ba
\rho + p &\geq & 0 \quad  \,\,\,\quad \Leftrightarrow  \quad \,\,\,\quad 
\psi - \frac{1}{2 \beta^3} \qty[\alpha^2 \partial_f \beta + \beta^2 \partial_r \alpha - 2 \alpha \beta \partial_r \beta ]\,\, \geq \,\, 0 \,,\\
p &\geq & 0
\quad  \,\,\, \quad \Leftrightarrow \quad \,\,\, \quad 
- \frac{1}{2  \beta^3}\qty[\alpha^2 \partial_f \beta  + \beta^2 \partial_r \alpha  - 2 \alpha \beta \partial_r \beta  + \frac{2}{r} \alpha \beta^2] \,\,\geq \,\, 0\,.
\ea
The first condition involves the kinematic quantity $\psi$, but can be phrased as a theory-dependent statement by making use of the  inverse of $\Omega$ with respect to $f$ applied to the equation of motion $\Omega(r,f) = 2M$, assuming it exists. For example, in curvature quasi-topological gravities with $r^3 h(\psi) = 2M$ according to~\eqref{eq:OmegaCurvature}, one may express $\psi$ in terms of the inverse $H$ of $h$ satisfying $H(h(x)) =x$, such that $H(S) = \psi$ with $S = \frac{2M}{r^3}$. See e.g.~\cite{Hennigar:2025yqm,Bueno:2026dln} for discussions relying on properties of such an inverse in higher-dimensional polynomial curvature quasi-topological gravities.

\section{Gravitational collapse into extremal regular black holes}\label{Sec:GravitationalCollapse}

In the following we will apply the previous discussions to describe the asymptotic formation of extremal regular black holes dynamically. The underlying interpetation is that any of the time-evolving spacetimes considered in this section are Vaidya solutions of general quasi-topological gravities with mass function $M(v)>0$. As a result, no energy conditions will be violated during such non-singular gravitational collapse processes, provided $\dot{M}(v) \geq 0$.

\subsection{Asymptotic formation of an extremal single-horizon regular  black hole}\label{SecSub:SingleHorizonFormation}

As a concrete example for a single-horizon extremal regular black hole, we will consider the extremal Hayward spacetime~\eqref{eq:f3HaywardExtremal}.
This metric can be obtained formally as a solution to the equations of motion of an integrable two-dimensional Horndeski theory~\cite{Boyanov:2025pes,Carballo-Rubio:2025ntd} and hence as a solution of a general quasi-topological gravity~\cite{Borissova:2026wmn,Borissova:2026krh} by inverting $f$ for $M$ and invoking~\eqref{eq:EqAlgebraic}, followed by the replacement $r \mapsto \varphi$ and $f \mapsto \chi$ offshell.
This inversion involves finding the roots of the following cubic polynomial in $M$,
\ba\label{eq:MPolynomial}
-\frac{32}{27} \psi r^2M^3 + 2 r^2 M - \psi r^5 &=& 0 \,.
\ea
In general, for any of the extremal regular black holes constructed in Sec.~\ref{Sec:StaticExtremalRBHs} starting from an ansatz~\eqref{eq:fk} and demanding that $M$ be the only dimensionful scale, such a reconstruction will involve finding the roots of higher-degree polynomials in $M$.
In the case of~\eqref{eq:MPolynomial}, there are two physical solution branches resulting in the generating functions
\ba\label{eq:OmegaPlusMinus}
\Omega_\pm(\varphi,\chi) &=& \frac{3 \varphi^2 \qty[\qty(1 \pm \imath \sqrt{3}) + \qty(1 \mp \imath \sqrt{3}) \qty[ \varphi^3 \psi^\frac{3}{2}+ \sqrt{\varphi^6 \psi^3-1}]^{\frac{2}{3}}]}{4 \qty[\varphi^9 \psi^3+ \varphi^6 \psi^{\frac{3}{2}} \sqrt{\varphi^6 \psi^3-1}]^{\frac{1}{3}}}\,,
\ea
with $\psi(\varphi,\chi) = (1-\chi)/\varphi^2$ offshell. The 
correct choice of branch producing the extremal positive-mass spacetime for a given $r \in (0,\infty)$ onshell depends on the location of its horizon $r_* = \frac{4 M}{3}$. One should choose $\Omega_+$ for $r\in (0,r_*]$ and $\Omega_-$ for $r \in [r_*,\infty)$. Accordingly, one may also anticipate two separate contributions $S_\pm$ to the action $S=S_+ + S_-$, which can be reconstructed explicitly at the two-dimensional level as described in~\cite{Boyanov:2025pes,Borissova:2026dlz}, and can be subsequently lifted to a four-dimensional quasi-topological gravity~\cite{Borissova:2026wmn,Borissova:2026krh} if desired.
Such a multi-valued characteristic function
 is reminiscent of the one required for spatially flat bouncing cosmologies in curvature quasi-topological theories with generating function of the form~\eqref{eq:OmegaCurvature}~\cite{Borissova:2026klg,Ling:2025ncw}.~\footnote{Such cosmological bouncing geometries satisfy $\psi = 0$ at some finite value of $h = \varrho_0$ in the ultraviolet regime, where $\psi = H^2$ is the squared Hubble parameter and $\varrho$ the matter energy density. Demanding simultaneously an infrared limit corresponding to general relativity with vanishing cosmological constant, in particular $h(0)=0$,  then requires that $h$  be multi-valued~\cite{Borissova:2026klg}. Hence also $h'(\psi_0) \to \pm\infty$ for some $\psi_0 > 0$, and accordingly $\beta \to \pm \infty$ at this point, cf.~equation~\eqref{eq:OmegaCurvature}. One may verify that the generating function combined out of the two branches~\eqref{eq:OmegaPlusMinus}, resulting in an extremal regular black hole, leads to a similar divergence $\beta = \partial_f \Omega \to \pm \infty$ for $f=0$, i.e., at the extremal horizon $r=r_*$. This is a physical divergence associated with the zero-temperature $T =  \frac{1}{4 \pi} \abs{\frac{\alpha(r_*,0)}{\beta(r_*,0)}}$~\cite{Borissova:2026rbi} limit of a hypothetic non-extremal regular black hole approaching extremality.}
It is also worth noticing that the above branches of the generating function do not contain any dimensionful constants. We are thus implicitly dealing with a highly non-perturbative gravitational theory free from additional scales  beyond the gravitational coupling $G_{\text{N}}$.~\footnote{This should be contrasted to the generating function obtained by inverting the metric function for the Hayward spacetime~\eqref{eq:f3Hayward} for $M$, which has a unique solution resulting in the generating function
	\ba
	\Omega(\varphi,\chi) &=& \frac{\varphi^3 \psi}{1 - \ell^2 \psi}\,.
	\ea
A single-horizon extremal regular black hole sourced from a theory with this particular generating function would be a fine-tuned configuration in which the mass $M$ just so happens to be equal to $\frac{3 \sqrt{3}}{4 }\ell$. However, any infusion of matter during a Vaidya collapse would perturb the spacetime away from extremality and into a non-extremal regular black hole.
} The result of coupling this implicitly identified general quasi-topological theory to a Vaidya source is a dynamical spacetime 
\ba\label{eq:f3HaywardExtremalv}
f(v,r) &=& 1 - \frac{2M(v) r^2}{r^3 + \frac{32}{27}M(v)^3}\,,
\ea
which can be used to describe the dynamical evolution of a single-horizon extremal Hayward regular black hole once it is formed. 
This function  is, however, not well-suited to describe the actual process of a gravitational collapse into an extremal regular black hole from an initially horizonless spacetime, as 
in the process from $M =0$ to $M >0$ arbitrarily small, the extremal horizon emerges instantaneously from the core.\\

\begin{figure}[t]
	\centering
	\includegraphics[width=0.75\textwidth]{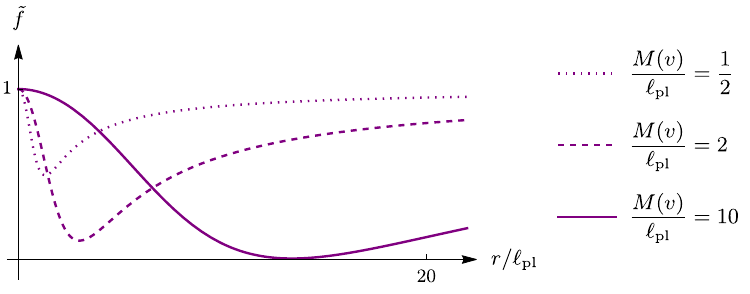}
	\caption{\label{Fig:HaywardExtremalDeformedCollapse} Formation of a single-horizon extremal regular black hole~\eqref{eq:f3HaywardExtremal} as a large-mass asymptotic limit of the deformed metric~\eqref{eq:f3HaywardExtremalvDeformed} with $\frac{L}{\ell_{\text{pl}}}=1$.
	}
\end{figure} 

Instead, we will consider here a Vaidya collapse describing the asymptotic formation of a single-horizon extremal regular black hole. To that end, we will deform the metric~\eqref{eq:f3HaywardExtremalv} away from extremality such that it is approached from  a horizonless compact object as a large-mass asymptotic limit. The gravitational  collapse will then in particular not produce a non-extremal regular black hole, i.e., classical mass inflation during the dynamical evolution and at its asymptotic limit is avoided.
This can be done by introducing a new length scale $L$ in addition to $M$, such that both 
scales contribute to singularity regularisation at $r=0$, while the metric function approaches the extremal one~\eqref{eq:f3HaywardExtremalv} asymptotically for large mass. One may abstractly view such a metric deformation as a deformation on the space of integrable two-dimensional Horndeski theories, and therefore as a mapping between an originally scale-free quasi-topological gravity and another one which involves an extra coupling constant, with the result that solutions are now horizonless spacetimes generically.
As an example, we can introduce the new length scale $L$ by deforming~\eqref{eq:f3HaywardExtremalv} into 
\ba\label{eq:f3HaywardExtremalvDeformed}
\tilde{f}(v,r) &=& 1-\frac{2 M(v)r^2}{r^3 + \frac{32}{27}M(v)^3 + 2 M(v) L^2}\,.
\ea
This deformation preserves the leading time-dependent Schwarzschild behavior at large $r$, while the effective regularisation length parameter at the core is now
\ba
\ell^2 &=&L^2 + \frac{16}{27}M(v)^2\,.
\ea
The length scale $L$ should be thought of as a non-dynamical ultraviolet scale which affects the spacetime at the core as soon as the gravitational collapse has started and $M$ attains a non-zero value.  In practice it prevents the exact formation of a black hole for finite $M$, as~\eqref{eq:f3HaywardExtremalvDeformed} does not have a positive root for $L > 0$. Instead, this function has a positive local minimum at $r = r_{\text{min}}$ with corresponding positive value of $\tilde{f}$ given in the limit of large $M$ by
\ba
r_{\text{min}}\,\,=\,\, r_*  + \mathcal{O}\qty(M^{-1})
%+ \frac{3}{4}  \frac{ L^2}{ M}+ \mathcal{O}\qty(M^{-3})
\,, \quad \quad \tilde{f}\qty(v,r_{\text{min}}) \,\,=\,\, \frac{9}{16 }  \frac{ L^2}{M^2}- \mathcal{O}\qty(M^{-4})\,,
\ea
where $r_* = \frac{4 M }{3}$ is the location of the would-be horizon of the would-be extremal spacetime~\eqref{eq:f3HaywardExtremalv} to which the deformed geometry asymptotes for large $M$, i.e., $r_{\text{min}} \simeq r_*$ in this limit. The deformed metric function itself as well as its radial derivative there are given by
\ba
\tilde{f}(v,r_*) 
\,\, = \,\,\frac{9}{16} \frac{L^2}{M^2} - \mathcal{O}\qty(M^{-4})\,, \quad \quad 
\tilde{f}'(v,r_*) 
\,\,=\,\, -\frac{27}{32}\frac{L^2}{M^3} + \mathcal{O}\qty(M^{-5})\,.
\ea
This 
provides a quantitative statement about the mass scale on which the deformed geometry approaches the one of a single-horizon extremal regular black hole~\eqref{eq:f3HaywardExtremalv}.

\begin{figure}[t]
	\centering
	\includegraphics[width=0.49\textwidth]{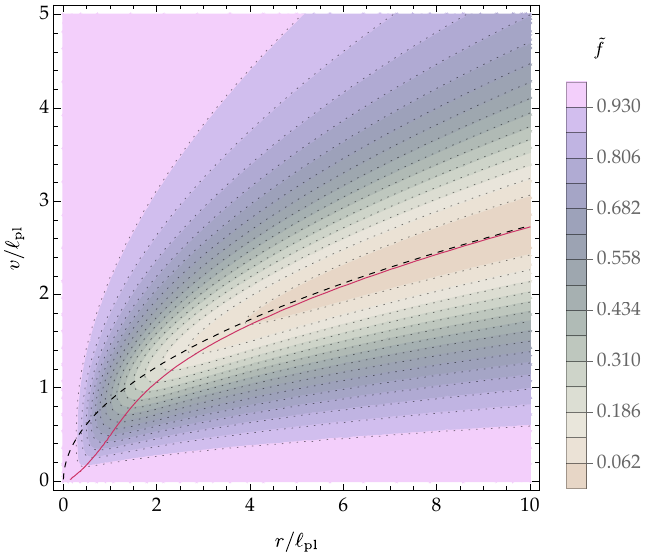}
	\includegraphics[width=0.485\textwidth]{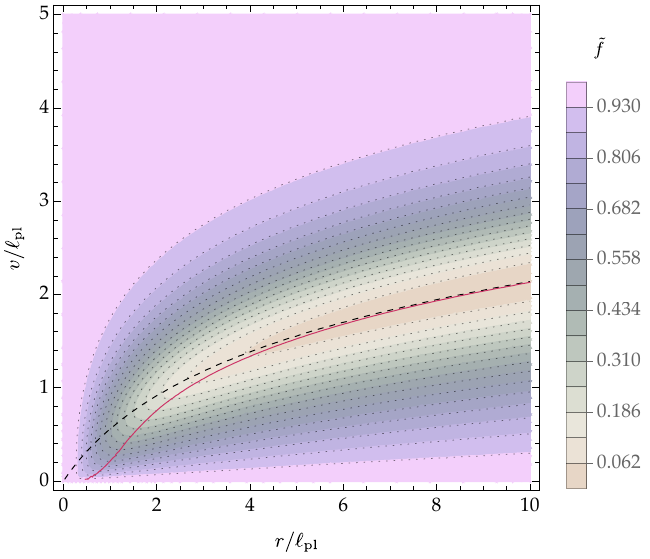}
	\caption{\label{Fig:HaywardExtremalDeformedCollapse2} Formation of a single-horizon extremal regular black hole~\eqref{eq:f3HaywardExtremal} as a late-time asymptotic limit of a Vaidya gravitational collapse modeled by the deformed metric~\eqref{eq:f3HaywardExtremalvDeformed} for a quadratic mass function $M(v) =  \frac{1}{\ell_{\text{pl}}}v^2$ in the left plot, and an exponentially growing mass function $M(v)= \ell_{\text{pl}} \qty(e^{\frac{v}{\ell_{\text{pl}}}}-1)$ in the right plot with $\frac{L}{\ell_{\text{pl}}}=1$. 
	The dashed black line marks the location $r_* =\frac{4M}{3} $ of the would-be horizon of the would-be extremal regular black hole geometry~\eqref{eq:f3HaywardExtremalv}. The solid red line is where $\tilde{f}'(v,r)=0$.
}
\end{figure}

Fig.~\ref{Fig:HaywardExtremalDeformedCollapse} shows the metric function~\eqref{eq:f3HaywardExtremalvDeformed} for increasing mass.
The timescale required to reach this extremal geometry approximately depends on the choice of mass function, cf.~Fig.~\ref{Fig:HaywardExtremalDeformedCollapse2} for an illustration comparing a quadratically and an exponentially growing mass function.

\subsection{Asymptotic formation of an inner-extremal double-horizon regular black hole}\label{SecSub:DoubleHorizonFormation}

We will now analogously consider the asymptotic formation of an inner-horizon extremal regular black hole from a horizonless object.
As a concrete example for a double-horizon extremal regular black hole, we will consider the   metric~\eqref{eq:f4Extremal} for $\varrho_- = \frac{1}{6}$, $\varrho_+ = 1$ and $\gamma=1$. Its time-dependent generalisation obtained by coupling an associated implicitly identified quasi-topological gravity to a Vaidya source is
\ba\label{eq:f4InnerExtremalv}
f(v,r) &=& 
\frac{\qty[r- \frac{1}{3} M(v) ]^3 \qty[r- 2 M(v) ]}{r^4 - M(v) r^3 + \frac{127}{54} M(v)^2 r^2 - \frac{19}{27}M(v)^3 r + \frac{2}{27} M(v)^4}\,,\,\,\,\,\quad
\ea
which can be used to describe an instantaneous formation and subsequent growth of an inner-extremal regular black hole. Similarly as before, such a function does, however, not allow us to model the gravitational collapse into an inner-extremal spacetime starting from a horizonless geometry.\\

\begin{figure}[t]
	\centering
	\includegraphics[width=0.75\textwidth]{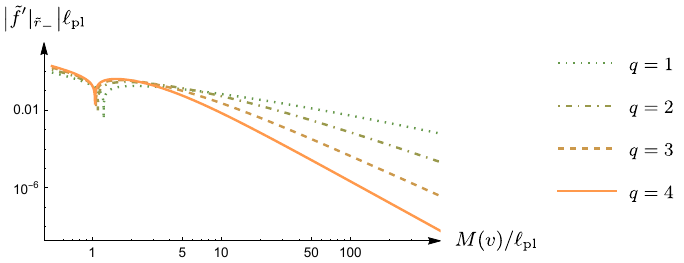}
	\caption{\label{Fig:InnerExtremalDeformedCollapsefPrimeqLogLog} Absolute value of the radial derivative of the deformed metric~\eqref{eq:f4InnerExtremalvDeformed} evaluated at the root $\tilde{r}_-$ 
		as a function of $\frac{M}{\ell_{\text{pl}}}$ for different $q$ with $\frac{L}{\ell_{\text{pl}}}=1$. The kink marks the minimum mass required for the existence of an inner horizon.
	}
\end{figure}

Instead, we will consider a Vaidya collapse describing the asymptotic formation of an inner-extremal regular black hole.
To that end, we introduce another length scale $L$ through the combination $M^{4-q} L^q$ for a positive integer $q$ smaller than or equal to $k=4$, and consider the following deformation of the metric~\eqref{eq:f4InnerExtremalv},
\ba\label{eq:f4InnerExtremalvDeformed}
\tilde{f}(v,r) &=& 
\frac{\qty[r- \frac{1}{3} M(v) ]^3 \qty[r- 2 M(v) ] + M(v)^{4-q} L^q}{r^4 - M(v) r^3 + \frac{127}{54} M(v)^2 r^2 - \frac{19}{27}M(v)^3 r + \frac{2}{27} M(v)^4 +M(v)^{4-q} L^q}\,.
\ea
The effective regularisation length parameter at the core is now
\ba
\ell^2 &=& 54 M(v)^{2-q} L^q + 4 M(v)^2\,.
\ea
As before, $L$ should be thought of as a non-dynamical ultraviolet scale which in practice prevents the exact formation of an inner-extremal horizon for finite $M$. 
The deformed metric for large $M$ has an inner horizon at $r = \tilde{r}_-$ with corresponding value of $\tilde{f}'$ given by
\ba\label{eq:rTildeMinus}
\tilde{r}_- \,\,=\,\,r_- + \mathcal{O}\qty(M^{1-q/3})\,,\quad \quad \tilde{f}'\qty(v,\tilde{r}_-)\,\,=\,\, - \frac{486\, 3^{\frac{2}{3}} 5^{\frac{1}{3}}}{37}  \qty(\frac{L}{M})^{2 q/3}\frac{1}{M} + \mathcal{O}\qty(M^{-q-1})\,,
\ea
where $r_- = \frac{M}{3}$ is the location of the inner horizon of the non-deformed geometry~\eqref{eq:f4InnerExtremalv}. While the leading-order asymptotics is $\tilde{r}_- \simeq r_-$ in the limit of large $M$, only for $q=4$ does the subleading term in the expansion of $\tilde{r}_-$ vanish for $M \to \infty$, while for $q=3$ it is constant, and otherwise it contributes to a subleading growth behavior of the inner horizon with mass.
 It is also important to emphasise that the leading term in the radial derivative above comes from evaluating it at the exact location of the deformed inner horizon $ \tilde{r}_-$ and expanding the result for large $M$. Fig.~\ref{Fig:InnerExtremalDeformedCollapsefPrimeqLogLog} shows a logarithmic plot of the absolute value of the radial derivative evaluated at $r = \tilde{r}_-$ as a function of $M$ for different $q$. This is the relevant quantity for determining how quickly the deformed geometry approaches extremality on its own, in the sense that $\tilde{f}'\qty(v,\tilde{r}_-) \to 0$ for large $M$, which is controlled by the exponent $q$.
A statement about how close asymptotically the deformed geometry comes to the original one~\eqref{eq:f4InnerExtremalv} can be made by evaluating the function~\eqref{eq:f4InnerExtremalvDeformed} and its radial derivative at the would-be inner horizon $r=r_-$ of the non-deformed inner-extremal spacetime~\eqref{eq:f4InnerExtremalv}, which gives
\ba
\tilde{f}\qty(v,r_-)
\,\,=\,\, \frac{486}{37 }  \frac{ L^q}{M^{q}} - \mathcal{O}\qty(M^{-2q})\, ,\quad \quad 
 \tilde{f}'(v,r_-) 
 \,\, =\,\, -\frac{160380}{1369}\frac{L^q}{M^{q+1}} + \mathcal{O}\qty(M^{-2 q -1})\,.\label{eq:fInnerExtremalDeformedPrime}
\ea
Notice, however, that $\tilde{r}_- = r_- $ never holds for finite $L$ and $M$, and therefore it is still the expression in~\eqref{eq:rTildeMinus} which matters in determining the mass scale of approximate extremality of the deformed geometry, whereas~\eqref{eq:fInnerExtremalDeformedPrime} serves as a cross check of the fact that the original metric~\eqref{eq:f4InnerExtremalv} with its exactly inner-extremal horizon at $r=r_-$ is recovered for $L\to 0$ at finite $M$, and is asymptotically approached as $M\to \infty$ at finite $L$.

\begin{figure}[t]
	\centering
	\includegraphics[width=0.75\textwidth]{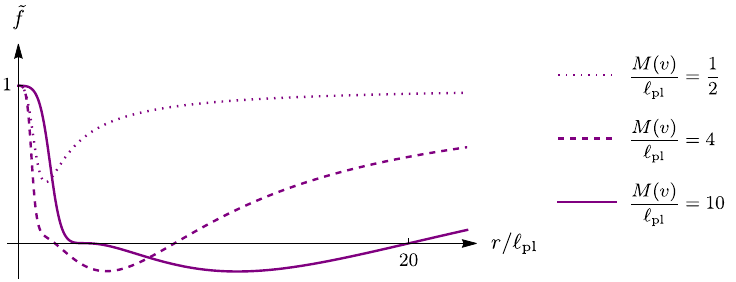}
	\caption{\label{Fig:InnerExtremalDeformedCollapse} Formation of a double-horizon inner-extremal regular black hole~\eqref{eq:f4Extremal} with $\varrho_- = \frac{1}{6}$, $\varrho_+ =1$ and $\gamma=1$ as a large-mass asymptotic limit of the deformed metric~\eqref{eq:f3HaywardExtremalvDeformed} with $q=4$ and $\frac{L}{\ell_{\text{pl}}}=1$.
	}
\end{figure}

\begin{figure}[t]
	\centering
	\includegraphics[width=0.56\textwidth]{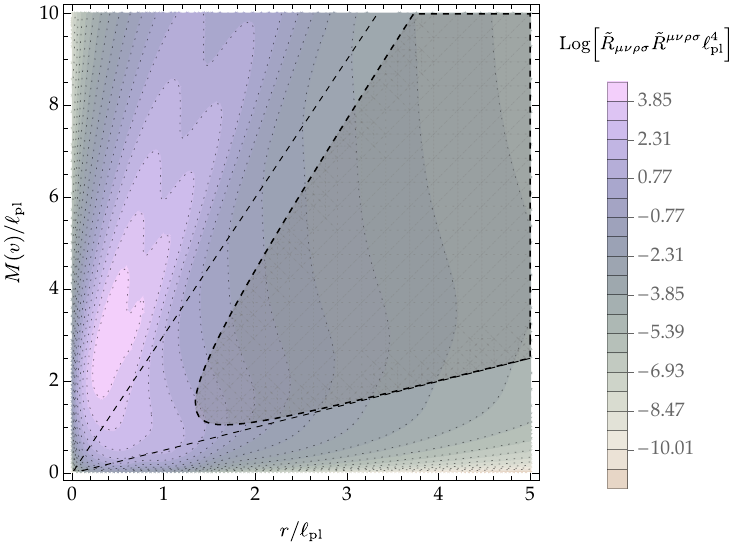}
	\caption{\label{Fig:InnerExtremalDeformedCollapseKretschmann} Kretschmann scalar of the deformed metric~\eqref{eq:f4InnerExtremalvDeformed} with $q=4$ as a function of $\frac{r}{\ell_{\text{pl}}}$ and $\frac{M}{\ell_{\text{pl}}}$ with $\frac{L}{\ell_{\text{pl}}}=1$. The shaded region is the trapped region where $\tilde{f}<0$. The dashed black lines emanating from the origin mark the locations $r_- = \frac{M}{3}$ and $r_+ = 2 M$ of the would-be inner and outer horizons of the original inner-extremal regular black hole geometry~\eqref{eq:f4InnerExtremalv}.
	}
\end{figure}

\begin{figure}[t]
	\centering
	\includegraphics[width=0.75\textwidth]{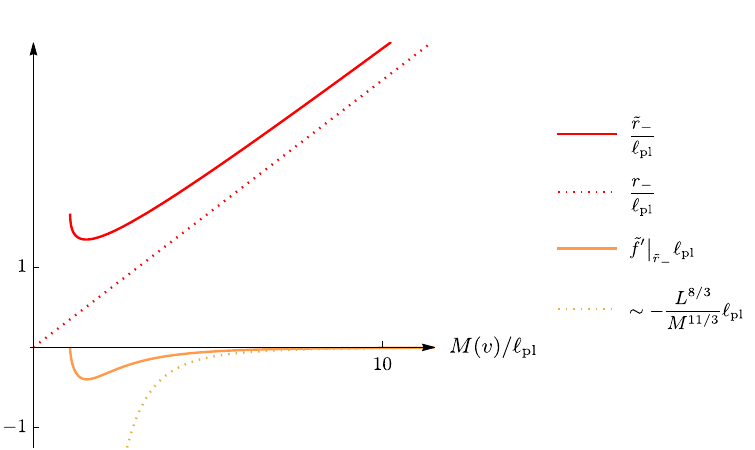}
	\caption{\label{Fig:InnerExtremalDeformedCollapsefPrime} Inner horizon $\tilde{r}_-$ of the deformed metric~\eqref{eq:f4InnerExtremalvDeformed} with $q=4$ compared to the inner horizon $r_-$ of the non-deformed metric~\eqref{eq:f4InnerExtremalv}, as well as radial derivative of the deformed metric~\eqref{eq:f4InnerExtremalvDeformed} evaluated at the exact location $\tilde{r}_-$ of its inner horizon compared to the asymptotic scaling~\eqref{eq:rTildeMinus}
		as functions of $\frac{M}{\ell_{\text{pl}}}$ with $\frac{L}{\ell_{\text{pl}}}=1$.
	}
\end{figure}

\begin{figure}[h!]
	\centering
	\includegraphics[width=0.49\textwidth]{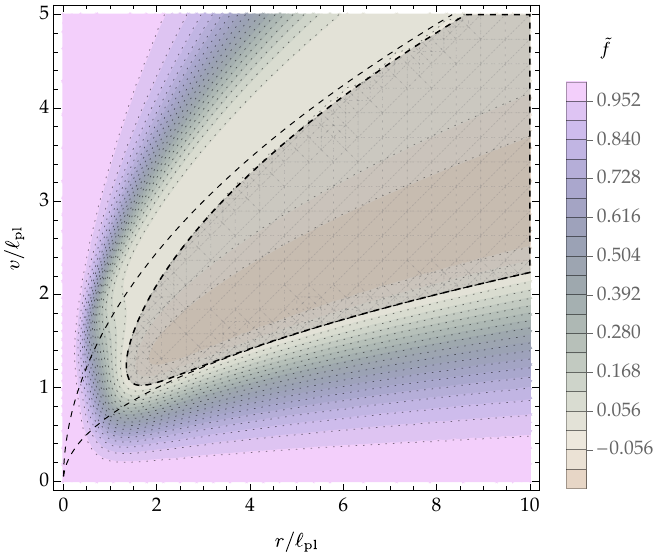}
	\includegraphics[width=0.485\textwidth]{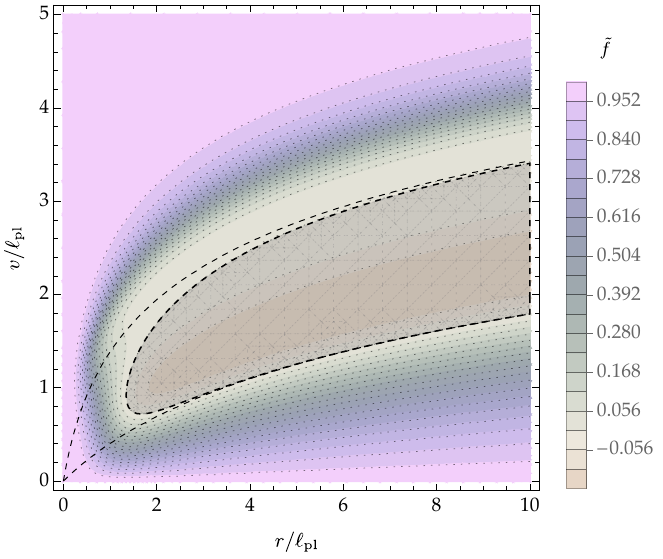}
	\caption{\label{Fig:InnerExtremalDeformedCollapse2} Formation of a double-horizon inner-extremal regular black hole~\eqref{eq:f4Extremal} with $\varrho_-= \frac{1}{6}$, $\varrho_+ =1$ and $\gamma=1$  as a late-time asymptotic limit of a Vaidya collapse modeled by the deformed metric~\eqref{eq:f4InnerExtremalvDeformed} with $q=4$ for a quadratic mass function $M(v) =  \frac{1}{\ell_{\text{pl}}} v^2$ in the left plot, and an exponentially growing mass function $M(v)=\ell_{\text{pl}} \qty(e^{\frac{v}{\ell_{\text{pl}}}}-1)$ in the right plot with $\frac{L}{\ell_{\text{pl}}}=1$. The shaded region is the trapped region where $\tilde{f}<0$. The dashed black lines emanating from the origin mark the locations $r_- = \frac{M}{3}$ and $r_+ = 2 M$ of the would-be inner and outer horizons of the original inner-extremal regular black hole geometry~\eqref{eq:f4InnerExtremalv}.}
\end{figure}

As a concrete example Fig.~\ref{Fig:InnerExtremalDeformedCollapse} shows the deformed metric function~\eqref{eq:f4InnerExtremalvDeformed} with $q=4$ for increasing mass. For completeness in Fig.~\ref{Fig:InnerExtremalDeformedCollapseKretschmann} we show the corresponding Kretschmann scalar  as a function of $r$ and $M$.
Fig.~\ref{Fig:InnerExtremalDeformedCollapsefPrime} shows the inner horizon $\tilde{r}_-$ of the deformed metric~\eqref{eq:f4InnerExtremalvDeformed} as well  as its radial derivative evaluated there and compared to the asymptotic scaling~\eqref{eq:rTildeMinus}, as functions of $M$.
The timescale required to reach an approximately extremal geometry, once a black hole has formed, depends on the choice of mass function, cf.~Fig.~\ref{Fig:InnerExtremalDeformedCollapse2} for an illustration comparing a quadratically and an exponentially growing mass function. 

\section{Discussion}\label{Sec:Conclusion}

We have constructed geometric models of static spherically symmetric and asymptotically flat single-horizon extremal and double-horizon inner-extremal regular black holes with a de Sitter core for generic black hole mass $M$ as the only dimensionful scale, which may be viewed as non-fine-tuned vacuum solutions to modified gravitational theories in four dimensions~\cite{Borissova:2026wmn,Borissova:2026krh}. The spherical reductions of these so-called general quasi-topological theories are by definition integrable two-dimensional Horndeski theories. As such, these gravitational theories admit exact Vaidya solutions~\cite{Boyanov:2025pes}. The constant mass $M$ of a static regular black hole vacuum solution then becomes a time-dependent function. We have used this effectively two-dimensional second-order dynamical framework to model the asymptotic formation of extremal regular black holes. No matter energy conditions are violated during these non-singular gravitational collapse processes when they are realised as quasi-topological Vaidya solutions rather than as generalised Vaidya solutions of general relativity~\cite{Husain:1995bf,Wang:1998qx} --- i.e., the violation of geometric convergence conditions~\cite{Borissova:2025msp,Borissova:2025hmj,Borissova:2026ohj} relevant to the singularity theorems~\cite{Penrose:1964wq,Hawking:1970zqf,Hawking:1973uf} and required for the defocusing of geodesics resulting in regular black holes is triggered by the modified gravitational dynamics rather than by the addition of exotic matter degrees of freedom. 

The subset of general quasi-topological gravities giving rise to the extremal regular black holes  discussed here will be free from additional scales beyond the gravitational coupling $G_{\text{N}}$. This is because the extremal spacetimes do not involve any dimensionful scales beyond the mass $M$, which arises as an integration constant when realising these spacetimes onshell.\\

This work explores the possibility of dynamical formation of regular black holes while elevating potential issues of mass inflation and associated classical instability in static and slowly evolving regular black holes with a non-extremal inner horizon~\cite{Brown:2011tv,Carballo-Rubio:2018pmi,Bertipagani:2020awe,Bonanno:2020fgp,Carballo-Rubio:2021bpr,DiFilippo:2022qkl,Bonanno:2022jjp,Bonanno:2022rvo,Carballo-Rubio:2022pzu,Bonanno:2023qhp,Carballo-Rubio:2024dca,Bonanno:2025bgc}. Thus, we have focused on the one hand on describing the asymptotic formation of a single-horizon extremal regular black hole without producing a non-extremal regular black hole in the dynamical process. In this scenario transient states of the gravitational collapse are described by a horizonless object which becomes more and more compact while approaching the geometry of a fully extremal regular black hole~\cite{DiFilippo:2024spj} as a candidate eternal endpoint which would be free from classical mass inflation~\cite{Carballo-Rubio:2022kad,Franzin:2022wai,DiFilippo:2024spj} and semi-classical inner-horizon instabilities~\cite{Barcelo:2020mjw,Barcelo:2022gii,McMaken:2023uue,Carballo-Rubio:2026gwg}, as well as instabilities due to Hawking evaporation~\cite{Hollands:2019whz,Balbinot:2023vcm,McMaken:2023tft,McMaken:2023uue,McMaken:2024tpc}, while the classical Aretakis instability of extremal regular black holes~\cite{Aretakis:2012ei} may  be avoided if the horizon is modeled to be of increasingly high degeneracy~\cite{Agrawal:2026oka}. Potential light ring instabilities~\cite{Cunha:2017qtt,Cunha:2022gde,Cardoso:2014sna,Franzin:2023slm} of these horizonless compact objects would have to be addressed.

\begin{figure}[t]
	\centering
	\includegraphics[width=0.6\textwidth]{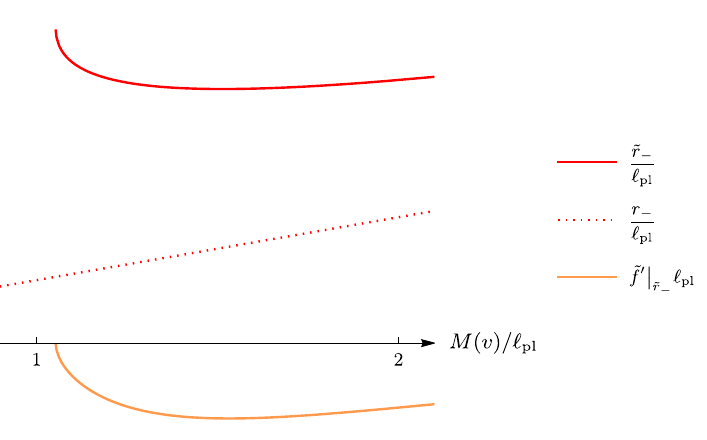}
	\caption{\label{Fig:InnerExtremalDeformedCollapse3} Zoomed-in plateau region of Fig.~\ref{Fig:InnerExtremalDeformedCollapsefPrime} for a small-mass black hole.
	}
\end{figure} 

In addition, we have considered the asymptotic formation of an inner-extremal regular black hole through a time-evolving regular black hole geometry which approaches inner extremality  
in inverse powers of $M$, multiplied by appropriate powers of a new fundamental length scale $L$, for large black hole mass.
This is also the scale which sets the time-dependent surface gravity $\kappa_-$ of the inner horizon~\cite{Carballo-Rubio:2024dca,Cropp:2013zxi}. Thus, naively, the timescale of mass inflation in these nearly-inner-extremal regular black holes, once they are of macroscopic size, would be 
proportional to positive powers of $M$.
In fact, since $\kappa_-(v) = \frac{1}{2} \eval{\tilde{f}'}_{\tilde{r}_-(v)}$ scales with inverse powers of the mass $M$ in the large-mass limit, while $M$ keeps increasing in time, the adiabatic condition $\abs{\derivative{ \kappa_-(v)}{v}} \ll \abs{\kappa_-(v)}^2$ assumed in the derivation of mass inflation in slowly evolving non-extremal geometries~\cite{Carballo-Rubio:2024dca} is easily violated in the large-mass regime of the considered examples. Hence, classical mass inflation could potentially be insignificant in these nearly-inner-extremal dynamical geometries. A similar conclusion may apply to the timescale of semi-classical instabilities of the inner horizon in these dynamical geometries, once thay have reached a certain degree of approximate extremality~\cite{Carballo-Rubio:2026gwg}. 

A potential drawback to the validity of these conclusions concerns the question of how such nearly-inner-extremal dynamical regular black holes can be realistically dynamically formed in the first place. 
The same time-dependent inner surface gravity function which would  render the timescales of the aforementioned instabilities large for a macroscopic black hole will initially increase in magnitude once a black hole of minimum mass $M_0 >0$ has formed, and will eventually reach a plateau for a small black hole with mass $M$ slightly above this threshold, cf.~e.g.~Fig~\ref{Fig:InnerExtremalDeformedCollapsefPrime} and Fig.~\ref{Fig:InnerExtremalDeformedCollapse3} which zooms into the relevant region.
The question is then whether mass inflation in such a regime would spoil the dynamical evolution towards an inner-extremal regular black hole,
or whether such a regime can be bypassed sufficiently rapidly in a realistic collapse.

Alternatively, one may at least kinematically envisage other qualitatively distinct possibilities of forming an inner-extremal regular black hole. One of these is a dynamical realisation of the scenario in Fig.~\ref{Fig:f4NonExtremal}, in which a horizonless compact object collapses into a multi-degenerate single-extremal horizon regular black hole, which subsequently evolves into a double-horizon inner-extremal regular black hole while maintaining extremality at all times. Such a scenario would avoid the question of mass inflation and may be subject to only mild semiclassical instabilities~\cite{Carballo-Rubio:2026gwg}, if it subsequently evaporates without leaving behind an eternal or long-lived inner-extremal trapped region. Another possibility is an evolving initially multi-horizon non-extremal regular black hole with an outer horizon and three separated inner horizons, or two degenerate inner horizons and another separated inner one, which eventually merge to produce a triple-degenerate inner horizon and hence a double-horizon inner-extremal regular black hole. There are, however, in general obstructions to realising such scenarios dynamically in the context of Vaidya solutions to modified gravitational theories considered here, without allowing for deformations of the metric which would introduce new length scales or resorting to the notion of approximate extremality. This obstruction lies in the fact that, as Vaidya solutions with all horizons required to be dynamical and $M$ the only dimensionful scale, the time evolution of the horizons is controlled by the mass function $M$ alone. Without further ingredients, such as taking into account perturbations, there is therefore  no dynamical mechanism in this framework which could trigger modifications of this evolution resulting in a separation or coalescence of horizons.
More refined models of gravitational collapse involving multiple matter scales or new gravity length scales may be required to produce such a scenario dynamically.

\begin{acknowledgments}
I would like to warmly thank the organisers and participants of the workshop {\it Crossing over the edge --- black hole interiors beyond general relativity}, which took place in July 2026 in Trieste, for delightful company and stimulating discussions on related topics. I am also grateful to Stefano Liberati for feedback on a previous version of the manuscript. This work is supported by STFC Consolidated Grant ST/X000575/1.
\end{acknowledgments}

\enlargethispage{20pt}

\bibliographystyle{jhep}
\bibliography{references}

\end{document}